\documentclass{article}
\usepackage{amsmath,amssymb}
\newcommand{\no}{\nonumber\\}
\newcommand{\be}{\begin{equation}}
\newcommand{\ee}{\end{equation}}
\newcommand{\ba}{\begin{eqnarray}}
\newcommand{\ea}{\end{eqnarray}}
\newcommand{\ci}[1]{\cite{#1}}

\newcommand{\la}[1]{\label{#1}}
\def\gl#1{(\ref{#1})}
\def\tr#1{\mbox{\rm tr}\left[#1\right]}
\def\Tr#1{\mbox{\rm Tr}\left[#1\right]}
\newcommand{\nonesep}{}
\newcommand{\mathd}{\mathrm{d}}
\newcommand{\bignone}{}
\newcommand{\tmop}[1]{\operatorname{#1}}

\begin{document}

\begin{center}
{\Large\bf Chiral dynamics from the hadronic string: general formalism}

\vspace{4mm}

A.~A.~Andrianov$^\dagger$,
    D.~Espriu$^\diamond$,
A.~Prats$^\diamond$\\
$^\dagger$ V.~A.~Fock Department of Theoretical Physics,
St. Petersburg State University, Russia\\
and\\
Istituto Nazionale di Fisica Nucleare, Sezione di Bologna, Italy,\\
E-mail: andrianov@bo.infn.it\\
$^\diamond$ Departament d'Estructura i Constituents de la Materia and CER for Astrophysics,
Particle Physics and Cosmology
Universitat de Barcelona, Spain\\
E-mail: espriu@ecm.ub.es;\, prats@ecm.ub.es
\end{center}

\begin{abstract}
QCD at long distances can be described by the chiral Lagrangian. On the other hand
there is overwhelming evidence
that QCD and all non-abelian theories admit an effective string description. Here we
review a derivation of the (intrinsic) parity-even
chiral Lagrangian by requiring that the propagation of the QCD string
takes place on a background where chiral symmetry is spontaneously broken.
Requiring conformal
invariance leads to the equation of motion of the chiral Lagrangian.
We then proceed to coupling the
string degrees of freedom to external gauge fields and we recover in this way
the covariant equations
of motion of the gauge-invariant chiral Lagrangian at ${\cal O}(p^2)$.
We consider next the parity-odd part
(Wess-Zumino-Witten) action and argue that this require the introduction of the spin degrees
of freedom (absent in the usual effective action treatment). We manage to reproduce the
Wess-Zumino-Witten term in 2D in an unambiguous way. In 4D the situation
is considerably more involved. We outline the modification of boundary
interaction that is necessary to induce the parity-odd part of the chiral Lagrangian.
\end{abstract}

\section{ Introduction: string propagation in a chirally broken
  background}

The history of attempts to describe the hadrons in the framework of
a string theory derived from, or at least inspired by, QCD encompasses already
more than 30 years (see, \cite{Ven}-\cite{solo} as well as
the reviews \cite{Rebbi}-\cite{polchin} and an incomplete
list of references therein).
The commonly cited arguments to justify a stringy description of QCD
 are the dominance of planar gluon diagrams
in the large $N$ limit \cite{largeN} `filling
in' a surface (interpreted as the world-sheet of a string),
the expansion in terms of surfaces built out of plaquettes
in strong-coupling lattice QCD \cite{lattice}, and to some extent
the incarnation of
Regge phenomenology \cite{regge} within QCD \cite{Lip}. Recently, the developments based
on the Maldacena conjecture \cite{Mald} and holographic duality \cite{dualstr} have added  further
strength to these arguments. The last but not the least is an advent of Nambu-Goto string
in lattice QCD of static heavy quark and antiquark \cite{Kuti}.

Clearly the simplest string models (bosonic string, supersymmetric string, ...) do
not lead to realistic amplitudes.
The paradigmatic example is the Veneziano amplitude\cite{Ven}; expanding it in powers of the
Mandelstam variables $s$ and $t$ one does not find  the right Adler zeroes and, of course,
reveals a tachyon in the spectrum. The supersymmetric version \cite{LS} partially solves
one of the problems by projecting out the tachyon, but the wrong chiral behavior persists.
Both difficulties are absent in the phenomenologically inspired Lovelace-Shapiro
amplitude \cite{lovelace}, but this amplitude does not seem to derive from any known string theory
and there are good reasons to believe that it is anyway incompatible with QCD 
asymptotics (see below).

Thus so far it is not yet clear what is a phenomenologically acceptable
QCD string action, even though there is a motivated
agreement based on universality considerations that in a certain
kinematic regime the  Nambu-Goto (or the Polyakov \cite{polya})
string action should be basically correct or, at least, provide the basic description.
A general characteristic of all the above amplitudes (including, incidentally the
Lovelace-Shapiro one) is that they lead to linearly rising trajectories.
General arguments
and recent work \cite{aaae} indicate that while this behavior corresponds to a
confining theory with
an infinite number of narrow resonances (in the large $N_c$ limit) it does not reproduce the
chiral properties of the QCD correlators. In fact, it can be proven that any
strictly linearly rising
Regge trajectory leads to complete degeneracy between the vector and the axial-vector
channels --- not
the way chiral symmetry is realized in QCD. Exponentially small
(of the form $\exp{(-an)}$, $n$ being
the principal quantum number) deviations are required and that means that none
of the existing
amplitudes can reproduce the chiral properties of QCD.

It is quite plausible that the main reason for the
presence of a tachyon in the spectrum and the wrong chiral
properties is a wrong choice
of the vacuum \cite{tachyon}. One can make a parallel with
scalar field theory  with the potential $V(\phi)=-\mu^2\phi^2 +
\lambda \phi^4$, that generates spontaneous symmetry breaking with a
sensible ground state, but where
perturbing around $\phi=0$ gives negative $m^2$ values
for all components. Thus we
assume that the string amplitudes obtained through the use
of the canonical vertex operators correspond to
amplitudes for excitations perturbed around the
wrong, unphysical vacuum.

A possible way to take into account the non-trivial nature
 of the QCD vacuum
was suggested in \cite{ADE} and developed in \cite{aabe}. Namely, one can assume
that in QCD chiral symmetry breaking takes place and
the light (massless in the chiral limit) pseudoscalar mesons form the background
of the QCD vacuum,
whereas other massive excitations are assembled into a string.
The massless pseudoscalars can be collected in a
unitary matrix $U(x)$. This matrix
transforms as
$U(x)\to U^\prime(x) = L U(x) R^\dagger$ under chiral transformations
belonging to $SU(3)_L\times SU(3)_R$  and describe excitations
around the non-perturbative vacuum. From the string point of view
$U(x)$ is nothing but a
bunch of couplings involving the string variable $x_\mu(\tau,\sigma)$.
The unitary matrix $U(x)$ has to be somehow coupled to the
boundary of the string, which is where flavor `lives'.
Our goal is to find eventually
a consistent string propagation in this non-perturbative background.

An essential property of string theory is, certainly, conformal invariance.
In the limit of large $N_c$ at least, the hadronic string action
should obey re-parameterization and conformal invariance as describing
zero-width, point-like resonance states\footnote{Perhaps
only in a dual manner -- after all there is a natural scale in QCD
and as we get to shorter and
shorter distances the partonic picture eventually sets in.}. Since conformal
invariance must hold when
perturbing the string around any vacuum we demand the new coupling to
chiral fields, living on the boundary, to preserve conformal invariance too
(compare with \cite{callan}). The requirement of conformal invariance will 
provide the equations of motion of the background fields and, indirectly, 
their Lagrangian.

In the present paper we begin by describing the basic characteristics of our approach.
We start by elucidating of how to incorporate 'quarks' at the end of the bosonic string,
in a manner respectful with conformal and the chiral symmetry properties of QCD and its vacuum,
by adding a suitable set of Grassmann variables \cite{ADE}, and, further on,
establish the general setting of the formalism
\cite{aabe}. We briefly review the results obtained in
this way, most notably the phenomenologically successful prediction of the ${\cal O}(p^4)$
equations of motion and the related low-energy
constants $L_1$, $L_2$ and $L_3$ of the effective chiral Lagrangian \cite{GL}.
We derive next a covariant
version of the results at order $p^2$ by coupling external gauge fields.
Then we proceed to the issue of deriving the odd intrinsic parity
part of the action from this approach and we immediately deduce the need of including the spin
degrees of freedom of the quarks (absent in the usual effective string treatment). By doing so
we obtain rather easily the anomalous part of the effective action in two
dimensions \cite{AEP}.  Finally we turn to
the four dimensional case that happens to be more involved. 
We discuss the general formalism and introduce a set
of operators (that eventually turn out to be embedded into an algebra) 
that implement the spin-flavor coupling.
In the subsequent sections
we derive the counterterms at the one and two loop level that are subjected to
vanish in order to guarantee conformal
invariance. We see at once that the 'quarks' represented
by the Grassmann variables at the end of the string cannot be in a $s=1/2$ angular 
momentum state if one
requires those counterterms to vanish and that they can be interpreted as
parts of equations of motion of local
non-linear sigma model. General
considerations regarding the algebra satisfied by the spin-flavor coupling operators
are presented.

\subsection{Basic concepts}\label{basicconcepts}

The hadronic string  is described by the following
conformal field theory action which has four dimensional Euclidean
space-time as target space
\begin{equation}
{\cal W}_{str}=\frac{1}{4\pi\alpha'}\int d^{2+\epsilon}\sigma
\left(\frac{\varphi}{\mu}\right)^{-\epsilon} \sqrt{|g|} g_{ik} \partial_i
x_\mu \partial_k x_\mu \ ,
\label{string}
\end{equation}
where,  for $\epsilon = 0$, in the conformal gauge $\sqrt{|g|} g_{ik} = \delta_{ik}$ and  one takes
$$x_\mu = x_\mu(\tau, \sigma);\quad
-\infty <\tau< \infty, 0< \sigma <\infty;\quad
i = \tau,\sigma \quad \mu=1,...,4 .$$
The conformal factor $\varphi(\tau, \sigma)$ is
introduced to restore the conformal
invariance in $2+\epsilon$ dimensions\footnote{Finally this factor becomes a dilaton
degree of freedom extending the four-component hadronic string to a five-component
one that however is beyond the scope of the present paper.}. The Regge trajectory slope
(related to the inverse string tension) is known
to be universal $\alpha' \simeq 0.9$ GeV$^{-2}$ from the meson phenomenology \cite{Anis}.

We would like to couple
the matrix in flavor space $U(x)$ containing  the meson fields in a chiral
invariant manner to the
string degrees of freedom while preserving
general covariance in the two dimensional coordinates and conformal
invariance under local scale transformations
of the two-dimensional metric tensor.
Since the string variable $x$ does not contain any flavor dependence,
we introduce two dimensionless Grassmann variables (`quarks')
living on the
boundary of the string sheet. The boundary quark fields
$\psi_L(\tau),\psi_R(\tau)$ transform in the fundamental representation
of the light-flavor group $SU(3)$.
 The subscripts $L,R$ are related to the
{\it chiral} spinors produced by the projectors (in what follows we use Euclidean space-time),
\be
P_{L} =\frac12(1 +\gamma_5),\quad P_{R} =\frac12(1 -\gamma_5),\quad
\gamma_5 =  \gamma_0 \gamma_1 \gamma_2 \gamma_3 \ .
\ee

A local hermitian action $S_b = \int d\tau L^{(f)}$ is introduced on the boundary
$ \sigma =0,\, -\infty < \tau < \infty$ to realize
the interaction with background chiral fields
$U(x(\tau)) = \exp(i \Pi(x)/f_\pi)$ where $f_\pi\simeq 90 MeV$,
the weak pion decay constant, relates the matrix field $\Pi(x)$ to a bunch of light
pseudoscalar mesons.

The boundary Lagrangian is chosen to be reparameterization invariant
and in its minimal form reads
\ba
L^{(f)}_{min}&=&\frac12 i \left(\bar\psi_L U (1 - z) \dot\psi_R  -
\dot{\bar\psi}_L U (1 +z)
\psi_R + \bar\psi_R U^+ (1 + z^*)
\dot\psi_L - \dot{\bar\psi}_R U^+ (1 - z^*) \psi_L\right)\ , \label{lagmin}
\ea
where a dot implies a $\tau$ derivative.
 The CP symmetry under transformation,
\be
\psi \longrightarrow \gamma_0 \psi; \quad \psi_R \longrightarrow
\gamma_0 \psi_L\ ;\quad
\psi_L \longrightarrow \gamma_0 \psi_R;\quad U \longrightarrow
U^\dagger\ , \label{CP}
\ee
requires purely imaginary constants $z = - z^* = \pm i|z|$.

It is easy to see that the string action \gl{string} is classically invariant under general
coordinate transformations of the two dimensional world sheet.
The fermion action is also automatically conformally invariant, because it does
not contain the two dimensional world sheet metric tensor since it can be
written as a line integral.

Upon obeying conformal invariance at the quantum level, one obtains the requirement
of a vanishing beta-function; in this case
a beta-functional of chiral fields and their derivatives as being coupling
constants of boundary string interaction. This beta-functional constraints the chiral field
$U(x)$  in order to have a consistent (i.e. conformally invariant) string propagation.
They have to be interpreted as the equations of motion for the collective field $U(x)$.
Adding the requirement of locality, the corresponding effective Lagrangian is
uniquely reconstructed.
In what concerns the parity-even part, this procedure will be explained in some more
details in section 2.

It is well known that the bosonic string is inconsistent at $d=4$ and that a dependence on
the conformal factor appears for non-critical dimensions. We regard this issue as collateral here, the
reason being that in a covariant treatment inconsistencies appear only at the one-loop level in
string perturbation theory. For an effective hadronic string of the type discussed here, this
would involve going beyond the $1/N_c$ limit and then the exact correspondence of QCD with an
string theory can be called into question anyway. In fact, nowhere 
in the calculation there appears any
interference between the requirement of conformal invariance 
for the $\Pi$-onic background and the string trace
anomaly. The matter deserves further study though.

\subsection{Feynman rules and perturbation theory}\label{feyrules}

In order to develop a perturbative expansion
we expand the function $U(x(\tau))$ in powers of the string coordinate
field $x_\mu(\tau) =x_{0,\mu} + \tilde x_\mu(\tau) $ around a constant $x_0$,
\begin{equation}
U(x(\tau)) = U(x_0) + \tilde x_\mu(\tau) \partial_\mu U(x_0) +
\frac12 \tilde x_\mu(\tau) \tilde x_\nu(\tau) \partial_\mu
\partial_\nu U(x_0) +\ldots \equiv  U(x_0) + {\cal V}(\tilde x)\ .
\label{expan}
\end{equation}
and look
for the potentially divergent
one particle irreducible diagrams.
We classify them according to the number of loops. Each additional
loop comes with a power of $\alpha^\prime$.
One can find a resemblance to the familiar
derivative expansion of chiral perturbation theory \cite{GL}.

The free fermion propagator is
\begin{equation}
\langle\psi_R (\tau) \bar\psi_L(\tau')\rangle = U^{\dagger} (x_0) \theta(\tau -
\tau')\ .
\end{equation}
If we impose $CP$ symmetry then
\begin{equation}
\langle\psi_L (\tau) \bar\psi_R(\tau')\rangle=
\langle\psi_R (\tau) \bar\psi_L(\tau')\rangle^\dagger = U (x_0) \theta(\tau -
\tau')\ ,
\end{equation}
for unitary chiral fields $U(x)$.

The free boson propagator projected on the boundary is
\begin{equation}
\langle \tilde x_\mu(\tau) \tilde x_\nu(\tau')\rangle = \delta_{\mu\nu}
\Delta (\tau -\tau')\ ,\quad
\Delta (\tau \rightarrow \tau') = \Delta (0) \sim
\frac{\alpha'}{\epsilon}, \quad \partial_\tau\Delta (\tau \rightarrow \tau') =
0\ ,
\end{equation}
the latter results hold in dimensional regularization (see below).

The normalization of the string propagator can be inferred from the
definition of the kernel of the N-point tachyon amplitude for the
open string\cite{Rebbi}.
\begin{equation}
\Delta (\tau_j -\tau_l) = - 2\alpha' \ln(|\tau_j -\tau_l|\mu)\ . \label{prop}
\end{equation}
Keeping in mind this definition let us determine the string
propagator in dimensional regularization, restricted on the boundary.
First we calculate the
momentum integral in $2+\epsilon$ dimensions,
\begin{equation}
\Delta_\epsilon (\tau) = \alpha' \Gamma\left(\frac{\epsilon}{2}\right) \
\left|\frac{\tau\mu\sqrt{\pi}}{\varphi}\right|^{-\epsilon}\ .
\label{epsprop}
\end{equation}
This dimensionally regularized propagator should be properly
normalized to reproduce (\ref{prop}). It can be done
by subtracting from (\ref{epsprop}) its value at $\tau\mu =1$
\begin{equation}
\Delta_\epsilon (\tau)|_{reg} = \alpha'
\Gamma\left(\frac{\epsilon}{2}\right)
\  \left\{
\left|\frac{\tau\mu\sqrt{\pi}}{\varphi}\right|^{-\epsilon}
- \left|\frac{\sqrt{\pi}}{\varphi}\right|^{-\epsilon}\right\}\ . \label{reg}
\end{equation}
Therefrom one unambiguously finds the relation
\begin{equation}
\Delta (0) =  - \alpha'
\Gamma\left(\frac{\epsilon}{2}\right)
\left|\frac{\sqrt{\pi}}{\varphi}\right|^{-\epsilon}
\stackrel{\epsilon \to 0}{=}
- 2\alpha'\left[\frac{1}{\epsilon} + C +
\ln\varphi\right] + {\cal O}(\epsilon) \equiv \Delta_\epsilon
- 2\alpha'\ln\varphi \ , \label{Dzero}
\end{equation}
where following the recipe of dimensional regularization
 we have taken $\epsilon < 0$ and hence
the first term in (\ref{reg}) vanishes at $\tau =0$.

The two-fermion, $N$-boson vertex operators are generated by the expansion
(\ref{expan}), from the generating functional
$ Z_b = \langle\exp(i S_b)\rangle$ and Eq.(\ref{lagmin}).
In particular, for the $L \to R$ transition one has vertices containing
$N$ derivatives of the chiral field and $N$ bosonic coordinates $\tilde x$
\begin{equation}
V = - \frac12 \left((1-z){\cal V}(\tilde x) \partial_\tau  +
(1+z)\partial_\tau \left[{\cal V}(\tilde x)\ldots\right]\right),
\label{nbos}
\end{equation}
and for the  $R \to L$ transition the Hermitian
conjugated vertex $V^+$ appears.

To implement the
renormalization process we perform a loop (equivalent to a
derivative) expansion
and proceed to determine the counterterms required
to make the theory finite. This will provide a
beta functional for the couplings $U(x)$ and their derivatives, which shall be
required to vanish up to
the two loop level in order to implement
the condition of vanishing conformal anomaly. The fact that we are working
with a boundary field theory makes
the required calculation quite manageable.
For a detailed derivation of the different Feynman diagrams
we refer the reader to \cite{aabe} and herein will report only
the final expressions.

\section{Summary of the parity-even sector results}\label{parityeven}

In this section we summarize the main results of \cite{aabe} which
are derived by using the previous Feynman rules.

At one-loop, the coefficient of the single pole gives the
appropriate counterterms. First we determine the fermion propagator counterterm at the one loop
order. Power counting indicates that this should be of ${\cal O}(p^2)$ in the chiral
expansion; i.e. two derivatives acting on the $U(x)$ field.
This gives the following counterterm
\begin{equation}
\delta^{(2)} U = \Delta(0) \left[\frac12 \partial^2_{\mu} U -
\frac{3 + z^2}{4}\partial_\mu U U^{\dagger}\partial_\mu U\right]\ , \label{resym}
\end{equation}
The coupling constant must be imaginary to provide the CP
symmetry. Its absolute value is determined by local
integrability, {\it i.e.} by the requirement that the equation $\delta U= 0$, derives from a
local Weinberg action,
\be
S_W = \int d^4x \frac{f_\pi^2}{4} \tr{\partial_\mu U \partial_\mu U^\dagger}\
;\quad \delta^{(2)} U = - \frac{\Delta(0)}{f_\pi^2} \frac{\delta S_W}{\delta
  U^\dagger (x)}\ . \label{weinb}
\ee
The latter one constraints  $z^2 = -1$. This condition also ensures the (perturbative) unitarity of
the chiral field.

The next step is to consider the renormalization of
the one-loop divergences in vertices with any number of
``bosons'' -- string coordinates  $x_\mu(\tau, \sigma = 0)$. Some
of the divergences are removed by the $U$ redefinition we just discussed,
as this automatically implies a counterterm for $\partial_\mu U$, the one-boson,
two-fermion tree-level vertex. This however is not sufficient to make
these vertices finite and an extension of the
boundary action is needed.

The relevant counterterms
can be parameterized\footnote{As compared to \cite{aabe} here we introduce the
{\it dimensionless} constants $g_i$ factorizing out the Regge slope scale
$\alpha'$ .} with three bare
constants $g_1$ , $g_2$ and $g_3$ (which are real if CP invariance holds),
\begin{eqnarray}
\Delta L_{bare}&= &\frac{i}{4} \alpha' (1 - z^2)\bar\psi_L\left(
(1 - z ) g \partial_\nu \dot{U} U^{\dagger}\partial_\nu U
- (1 + z ) g^* \partial_\nu U U^{\dagger}\partial_\nu \dot{U} \right.\nonumber\\
&&\left.+  z g_3 \partial_\nu U U^{\dagger} \dot{U} U^{\dagger} \partial_\nu U
\right)\psi_R  + \mbox{\rm h.c.} \ , \label{count}
\end{eqnarray}
where the complex constant $g$ is related to real constants from
\cite{aabe} as follows,
\be
g_1 = 2 (\mbox{\rm Re} g + \mbox{\rm Im} g );\quad
g_2 = 2 (\mbox{\rm Re} g - \mbox{\rm Im} g )\ .
\ee
This definition will be justified after generalization of boundary action
in Sec. 5.
Renormalization is accomplished by redefining the couplings $g_i$
in the following way
\begin{equation}
g_i = g_{i,r} - \frac{\Delta(0)}{\alpha'}. \label{gren}
\end{equation}
In spite of the fact that
the new vertices are higher-dimensional , it turns out \ci{aabe} that
their contribution into the trace of the energy-momentum tensor
vanishes once the requirements of CP invariance and unitarity of $U$ are
taken into account
and therefore conformal invariance is not broken (see \cite{string}).
One can  also prove \cite{aabe} that vertices with more boson legs are made
automatically finite once the renormalization of $U$ and $g_i$ has been
performed --- this completes the renormalization program
at the one-loop level.

At two loops calculations are certainly more involved, but still relatively simple
since we are working in a boundary field theory. We need to consider here
only the renormalization of the fermion propagator. At this order there
are several contributions: genuine two-loop contributions, one-loop $U$-counterterms
inserted in one-loop diagrams (both in the propagator and in the vertices), and
also the counterterms we just discussed (\ref{count}) inserted in the vertices
of one-loop diagrams.
We do not provide detailed formulae here because in section 5 we analyze in detail
a generalization of these results that take into account spin effects. The expressions that
are relevant for this section can be obtained from those in section 5 by
simplifying
to the present case.

After interpreting the vanishing beta-function condition as an equation
of motion of the (parity-even) chiral Lagrangian one unambiguously obtains
the low-energy constants \cite{GL} appearing in the order $p^4$ chiral Lagrangian.
They are expressed
in terms of the product of the Regge trajectory slope $\alpha'
\simeq 0.9$ GeV$^{-2}$,  $f_\pi^2$ and certain rational numbers
(equivalently they can be characterized by the ratio of  $f_\pi^2$ to
the hadron string tension $T = 1/ 2\pi \alpha'$). The unique solution is
\ba
&&g_{1,r} = - g_{2,r} =  - g_{3,r} = 2 ;\quad g = i;\\
&&2 L_1 = L_2 = -\frac12 L_3 = \frac{ f_\pi^2\alpha' }{8}
= \frac{ f_\pi^2}{16\pi T} \simeq 10^{-3} \equiv \xi.
\label{dim4}
\ea
This prediction  fits quite well the
phenomenological values \cite{expt}. The small dimensionless parameter $\xi$
is an expansion parameter of Chiral Perturbation Theory and represents a
natural scale for dimension-4 structural constants. Respectively in Section 5
it will be used in the parameterization of a most general local chiral
Lagrangian.

Meantime the Lagrangian (\ref{lagmin}) only
contains intrinsic parity-even terms
 and
does not contain any operators which can eventually
entail the anomalous P-odd part of the chiral dynamics. We turn to
this interesting question next in sections 4  and 5.

\section{Covariant equations of motion}

Let us incorporate external abelian gauge fields
into the boundary chiral action \gl{lagmin}. The tree-level
Lagrangian has to be
translation- and time-reparameterization invariant and invariant under the
gauge transformation, generated by an electric charge $Q$,
\begin{eqnarray}
\label{basicL}
\psi(x) \Longrightarrow e^{i\Lambda(x) Q} \psi(x)\,
,\quad A_\mu(x)\Longrightarrow A_\mu(x)+ Q \partial_\mu\Lambda(x)\,,
\nonumber
\end{eqnarray}
\begin{equation}
U(x) \Longrightarrow  e^{i\Lambda(x) Q} U(x) e^{-i\Lambda(x) Q}.
\end{equation}

Thus, in principle,
the boundary Lagrangian can be constructed with the help of the
covariant derivative projected on the boundary,
\be
\dot x_\mu \left(\partial_\mu -i  A_\mu(x)\right) =
\partial_\tau - i \dot x_\mu A_\mu\, \Longrightarrow  e^{i\Lambda(x)
  Q} \left(
\partial_\tau - i \dot x_\mu A_\mu\right) e^{-i\Lambda(x) Q}.
\ee
\begin{equation}
{\cal L} = \frac{i}{2}  \bar\psi_L\left\{(1-z) U(x)
(\partial_\tau - i \dot x_\mu A_\mu)
+ (1+z)(\partial_\tau - i \dot x_\mu  A_\mu) U(x)\right\} \psi_R
+ h.c.
\end{equation}

However it turns out that for such a Lagrangian the corresponding fermion propagator is
not gauge invariant, rather being
bilocally covariant. As a consequence, the divergences do not form a
gauge covariant combination and one ends up with equations of motion 
that do not derive from a manifestly gauge invariant Lagrangian. 
One has to proceed to the  fermion
fields dressed with the Dirac string phase factor so that the Lagrangian
is written as
\begin{equation}
{\cal L} = \frac{i}{2}  \bar\Psi_L\left\{(1-z)\widetilde U(x)
(\partial_\tau - i \dot x_\mu \Delta A^\bot_\mu)
+ (1+z)(\partial_\tau - i \dot x_\mu \Delta A^\bot_\mu)\widetilde U(x)\right\} \Psi_R
+ h.c.
\end{equation}
Herein we redistribute 
the e.m. interaction between dressed 
fermions ($\Psi=e^{-i\varphi(x)_\parallel Q}\psi$), chiral fields
\begin{equation}
U(x) \rightarrow \widetilde U(x) = e^{-i\varphi(x)_\parallel Q} U(x) e^{i\varphi(x)_\parallel Q};
\end{equation}
and the covariant derivative. $\varphi_\parallel$ is defined as
\begin{equation}
\varphi_{\parallel} = \widetilde x_{\mu}(\tau)A_\mu(x_{0}) + \sum_{n=1}^{\infty}\frac{1}{(n+1)!}
\widetilde x_{\mu}(\tau)\widetilde x_{\nu_{1}}(\tau)\cdots\widetilde x_{\nu_{n}}(\tau)
\partial_{\nu_{1}}\cdots\partial_{\nu_{n}} A_\mu(x_{0});
\end{equation}
whereas the remaining transversal part of the covariant derivative reads
\begin{equation}
\Delta A^\bot_\mu =
\sum_{n=1}^{\infty}\frac{n}{(n+1)!}
\widetilde x_{\nu_{1}}(\tau^\prime)\cdots\widetilde x_{\nu_{n}}(\tau^\prime)
\partial_{\nu_{1}}\cdots\partial_{\nu_{n-1}}F_{\nu_{n}\mu}(x_{0}) .
\end{equation}

Now in order to control the conformal symmetry we expand $\widetilde U(x)$ around a
constant background $x_0$
and look for the potentially divergent, one particle irreducible
diagram:
\begin{eqnarray}
\widetilde U(x) &&\!= (1 - i\varphi_\parallel - \frac{1}{2}\varphi_\parallel^2 +
\ + \cdots ) (U(x_0) + \widetilde
x_\mu\partial_\mu U(x_0) + \frac{1}{2}\widetilde x_\mu \widetilde
x_\nu\partial_\mu \partial_\nu U(x_0) + \cdots )\nonumber\\
&&\times (1 + i\varphi_\parallel -
\frac{1}{2}\varphi_\parallel^2 + \cdots )
\nonumber\\
&& = U(x_0) +\widetilde x_\mu D_\mu U(x_0) + \frac{1}{2}\widetilde x_\mu
\widetilde x_\nu D_\mu D_\nu U(x_0)+ \cdots  
 \label{emvert}
\end{eqnarray}
where the covariant derivative acts in the adjoint representation
$$  D_\mu U \equiv [ D_\mu, U] = \partial_\mu U - i [A_\mu, U] .
$$
Using the perturbation expansion and  vertex operators from
Eq.~\gl{emvert}
one arrives
 to the following expression:
$$
\frac{1}{4} U(x_0)^\dag\ D_\mu U(x_0)\ U(x_0)^\dag\ D_\mu U(x_0)\ U(x_0)^\dag
\{(3+z^2) \Delta (0) + (1-z^2) \Delta (\tau_A -\tau_B) \}
\Theta(\tau_A -\tau_B)$$
\begin{equation}
-\frac{1}{2} \Delta (0)\Theta(\tau_A -\tau_B) U(x_0)^\dag\ D_\mu D_\mu U(x_0)\
 U(x_0)^\dag .
\end{equation}
The divergence is eliminated by introducing an appropriate counterterm
\begin{equation}
\delta U = \frac12 \Delta (0)
\left\{D_\mu^2 U - \left( \frac{3+z^2}{2}\right) D_\mu U \
U^\dag\  D_\mu U \right\}
= 0,
\end{equation}
and conformal symmetry is restored if $\delta U$ vanishes as before.
The value $z^2 = -1$ provides the integrability of this chiral
dynamics,
i.e. its origin from the local gauged Weinberg Lagrangian.
We see that the dressed field Lagrangian produces the gauge invariant
chiral dynamics which is determined unambiguously at one-loop level.
The extension of the covariantization to the $p^4$ terms is in progress.

\section{Two-dimensional QCD and the WZW term}

The chiral bosonization of hadronic string presented in previous sections
is certainly incomplete as it does not include any quark spin degrees of
freedom and therefore does not generate parity-odd chiral dynamics in
the
form of the chiral anomaly in the equations of motion and
the Wess-Zumino-Witten chiral Lagrangian. To understand  the way 
parity-odd terms could emerge from the hadronic string built over the 
chirally broken QCD vacuum
we investigate the toy model of two-dimensional QCD.

While a parity-even chiral-field interaction on the line may be
qualitatively associated with
vector quark currents a parity-odd interaction must have relation to
axial-vector currents.  However in QCD$_2$, in fact, vector and axial-vector
fields couple to quarks with the same matrix vertex. Indeed,
 in two (Euclidean) dimensions  the structure of Dirac $\gamma$ matrices (in
 terms of the Pauli matrices $\sigma_a$),
$$\gamma_0 = \sigma_1;\, \gamma_1 = \sigma_2, \, \gamma_2\,
(\mbox{analog of}\, "\gamma_5") = \sigma_3 = -i\gamma_0 \gamma_1,$$
allows to relate axial-vector and vector vertices as follows,
\be
\gamma_\mu \gamma_2 = i \epsilon_{\mu\nu}\gamma_\nu,
\ee
in terms of antisymmetric tensor  $\epsilon_{\mu\nu} = -
\epsilon_{\nu\mu},\,  \epsilon_{01} = 1$.
Meantime
the O(2) algebra is generated by
$\sigma_{\mu\nu} \equiv -\frac12 i [\gamma_\mu,\gamma_\nu] = \epsilon_{\mu\nu} \gamma_2$.

Accordingly the boundary Lagrangian may equally well include two types of couplings,
\begin{eqnarray}
L^{(f)} &\equiv& \frac12 i  \left\{\bar\psi_L \left[ \{\partial_\tau, U\}  +
\widehat{F}_{\mu\nu} \dot x_\mu
   \partial_{\nu} U\right] \psi_R  + \bar\psi_R \left[ \{\partial_\tau,
  U^{\dagger}\} - \widehat{F}^\dag_{\mu\nu} \dot x_\mu
   \partial_{\nu} U^{\dagger}\right]\psi_L \right\};\nonumber\\
   \widehat{F}_{\mu\nu} &\equiv& z \delta_{\mu\nu} +
i g_A \epsilon_{\mu\nu}. \label{lagdim2}
\end{eqnarray}
 The
CP symmetry \gl{CP} of the Lagrangian \gl{lagdim2} holds only if
\be
z = - z^*;\quad g_A =  g_A^*;\quad \widehat{F}_{\mu\nu} =
- \widehat{F}^\dag_{\mu\nu}. \label{CPfull}
\ee

Now we develop string perturbation theory expanding the function $U(x)$ in powers of the
string coordinate
field $x_\mu(\tau) =x_{0,\mu} + \tilde x_\mu(\tau) $, then expanding the boundary action
in powers of $\tilde x_\mu(\tau)$
and finally looking
for divergences, i.e. violations of conformal symmetry. 
At one loop one obtains the following condition
 to preserve conformal symmetry ,
\begin{equation}
- \partial_\mu^2 U +\frac12(3+z^2 - g^2_A)\partial_\mu U U^\dagger \partial_\mu U
- i g_A \epsilon_{\mu\nu}\partial_\mu U U^\dagger \partial_\nu U = 0.
\label{anoma}
\end{equation}
Unitarity of chiral fields and
local integrability of Eqs. of Motion constrains the coupling constants to
fulfill the relation $g^2_A - z^2= 1$. The naive QCD value (if we trust the
arguments presented in Appendix A) is $g_A=1$. This choice ($z=0, g_A = 1$) corresponds 
to the correct value\cite{bosqcd} of the
dim-2 anomaly (last term in (\ref{anoma})). Thus in QCD$_2$ the hadron 
string induces the WZW action
from the vanishing the boundary $\beta$ function already at one-loop level.

In turn, in QCD$_4$ the anomaly and the WZW action have 
dimension 4 and 5 respectively and therefore they
are generated by cancellation of  two-loop divergences. Therefore the  antisymmetric tensor
$\epsilon_{\mu\nu\rho\lambda}$ in anomalies must arise from the
algebra of $\widehat{F}_{\mu\nu}$ matrices.

On the other hand, the boundary quark fields
$\psi_L(\tau),\psi_R(\tau)$
are, in fact, one-dimensional and
one should not expect that they
realize the fundamental, spin-1/2 representation of the Poincare
group. This is because the projection on a line is not uniquely
defined (see Appendix A) and to correct this projection consistently with
the conformal symmetry and integrability we eventually have to introduce a
more
complicated algebra than the conventional Clifford one.
Certain arguments in favor of
this extension will be given in the next Section.

\section{General formalism: renormalization at the one- and two-loop order}

In the next sections we are going to translate the ideas above to
the four dimensional case, develop the equations up to two loops, and
try to set a general framework for the search of solutions satisfying
unitarity of the $U$ matrices and CP invariance. The goal of this
section is to ensure the renormalizability at one- and two-loops of our model.

\subsection{One loop fermion propagator: A first guess.}
The starting point in this section is the following Lagrangian on
the boundary for the fermions $\psi_L$ and $\psi_R$, analogous to
Eq. (\ref{lagdim2}).

\begin{eqnarray}
L^{(f)}&=& \frac12 i  \left\{\bar\psi_L \left[ \{\partial_\tau,
U\}  + {F}_{\mu\nu} \dot x_\mu
   \partial_{\nu} U\right] \psi_R  + \bar\psi_R \left[
  \{\partial_\tau, U^{\dagger}\} - {F}^{\dagger}_{\mu\nu} \dot
  x_\mu
   \partial_{\nu} U^{\dagger}\right]\psi_L \right\};\no
F_{\mu\nu}&=&z\delta_{\mu\nu}+g_\sigma \sigma_{\mu\nu}.\label{Lagran4D}
\end{eqnarray}

The interaction term proportional to the $F_{\mu\nu}$ encodes the spin
degrees of freedom of the fermion variables once projected to the line
(boundary of the string) where the fermions live. This is why, as a
first choice, we have included $\sigma_{\mu\nu}$ in analogy to its
two-dimensional partner $\widehat F_{\mu\nu}$.

Following the procedure explained in Sec.\ref{parityeven} for the parity
even part, and following the rules of Sec. \ref{basicconcepts} and
\ref{feyrules} we will expand the $U(x(\tau))$ in powers of the
coordinate fields $x_\mu(\tau)=x_{0,\mu}+\tilde x_\mu(\tau)$. This will
bring us a variety of operators; some vertices will
contain $F_{\mu\nu}$ (those coming from the expansion of the second
term in the Lagrangian) and some will not (those coming from the first
term in the Lagrangian). We will perform all computations with this
expanded Lagrangian.

After computation of the diagrams we see that the divergence in
the $R\to L$ part of the fermion propagator\footnote{From now on we focus our analysis on
the divergences in the  $R\to L$ part of the fermion propagator having in mind
that the  $L\to R$ part is reproduced by means of hermitian conjugation.} 
at one loop takes the form
\begin{equation}
  O_1=\frac{1}{4} \theta ( A - B )\Delta(0)[ - 2 \partial^2 U + (
   3 \delta_{\alpha \beta} + F_{\beta \alpha} - F_{\alpha \beta} +
   F_{\alpha\mu} F_{\beta \mu} )\partial_{\alpha} U \nonesep U^{\dag}
   \partial_{\beta} U ].
\end{equation}
We can compare this equation with Eq.(\ref{resym}). 
Borrowing the ideas from the two-dimensional case and motivated
by the discussion in Appendix A we write (\ref{Lagran4D}).

Applying this definition to the one loop propagator we see that there
appear two different channels. One channel related with the trace of
the $O_1$, and another channel defined by its traceless part.
Let us compute them separately.

On one side we can perform a trace in spinor space and find
\begin{equation}
\frac{1}{2}\tr{O_1}=\frac{1}{4} \theta ( A - B ) [ - 2
   \partial^2 U + ( 3 + z^2 +3 g_\sigma^2 )
   \partial_{\mu} U \nonesep U^{\dag} \partial_{\mu} U ]
\label{scalarchann}
\end{equation}
where the identity
\begin{equation}
\sigma_{\mu\rho}\sigma_{\nu\rho}=3 \mathbb{I}
\delta_{\mu\nu}-2i\sigma_{\mu\nu}
\end{equation}
has been used.
We can recover the already known result of \cite{aabe} by making
$g_\sigma=0$.

The divergence in Eq.(\ref{scalarchann}) is eliminated by introducing
an appropriate counterterm $U\to U+\delta U$
\begin{equation}
\delta U=\Delta (0)\bigg [\frac{1}{2}\partial_\mu^2 U
  -\frac{3+z^2+3g_\sigma^2} {4} \partial_\mu U U^{\dagger} \partial_\nu U
  \bigg ],
\end{equation}

 Conformal symmetry is restored (the $\beta$-function is zero) if the
 above contribution vanishes. The unitarity of $U$ is compatible with
 the conformal symmetry saturation condition only if we demand
\begin{equation}
z^2+3g_\sigma^2=-1.
\end{equation}
When taking into account the CP symmetry condition $z=-z^*$,
$g_\sigma=-g_\sigma^*$ one find the following bounds on these couplings
constants
\begin{eqnarray}
&&0 \leq |z| \leq 1,\no
&&\frac{1}{\sqrt{3}} \geq |g_\sigma| \geq 0, \label{bounds}
\end{eqnarray}
and if the ein-bein projector gives a correct hint (see Appendix
A) then $|z|=|g_\sigma| = 1/2$.

Let's explore now the other channel, i.e. the traceless part of
$\delta U$. The latter is, in this case, the
part proportional to $\sigma_{\mu\nu}$, thus
\begin{equation}
\bar {O}_1 = \theta(A-B)i\frac12 \Delta(0)\left\{-i g_\sigma -
g_\sigma^2\right\} \sigma^{\mu\nu}\partial_\mu U
U^{\dagger}\partial_\nu U
\end{equation}
This is a completely new term not observed before which comes directly
from the inner space in $F_{\mu\nu}$.

We remark that for the $L \to R$ propagation the divergence is just
complex conjugated, i.e. has the coefficient $\left\{i g^*_\sigma +
(g_\sigma^*)^2\right\}$. This fixes $ g_\sigma = 0,-i$ in order to
make zero the non-scalar part. Both choices seem to be unacceptable
because for $ g_\sigma = 0$ one does not reproduce the Wess-Zumino
action and for $ g_\sigma = -i$ one cannot provide the vanishing
$\beta$-function in the scalar channel for the CP invariant choice
(\ref{CPfull}). After this negative result we must accept that this,
most intuitive, choice of $F_{\mu\nu}$ is not convenient for our
purposes.

\subsection{One loop fermion propagator: General form.}

At one loop we have been already able to see the incompatibility of
the guess (\ref{Lagran4D}) with the unitarity and CP conditions for the
model. At this point we must generalize our strategy
allowing for more general forms of $F_{\mu\nu}$. This will make the
spin content not well defined since general $F_{\mu\nu}$ can follow a
more complicated algebra than the Clifford one. Exactly as in the
previous guess, here we must consider that the $F_{\mu\nu}$ acts on an
internal spinor space. This requires some care in the 
computations in order to keep the right ordering of the
$F_{\mu\nu}$'s. This will be crucial in the renormalization process.

In this framework we recover the original Lagrangian (\ref{Lagran4D})
leaving $F_{\mu\nu}$ unspecified for the time being. 
The complete one loop contribution
to the fermion propagator in its general form reads.
\begin{eqnarray}
&&\frac{1}{2} \theta ( A - B ) \Delta ( 0 )\{-U^{\dag}
    \partial^2 U \nonesep U^{\dag}+\frac{1}{2}U^{\dag}
    \partial_{\sigma} U \nonesep U^{\dag} \partial_{\lambda} U
    \nonesep U^{\dag}[ 3 \delta_{\sigma \lambda} - ( F_{\sigma
    \lambda} - F_{\lambda \sigma} ) + F_{\sigma \gamma} F_{\lambda
    \gamma} ]\}\nonumber\\
&&+\frac{1}{4}\theta ( A - B ) \Delta ( A - B )U^{\dag}
    \partial_{\sigma} U \nonesep U^{\dag} \partial_{\lambda} U \nonesep
    U^{\dag}[ \delta_{\sigma \lambda} + ( F_{\sigma \lambda} -
    F_{\lambda \sigma} ) - F_{\sigma \gamma} F_{\lambda \gamma} ]
\end{eqnarray}
From this we can identify the general condition that unitarity of the
$U$ imposes,
\begin{equation}
\delta_{\sigma \lambda} - F_{\sigma \lambda} + F_{\lambda \sigma} +
 F_{\sigma \gamma} F_{\lambda \gamma}=0.\label{algebra1}
\end{equation}
This relation is a
first constraint for the algebra we have alluded to.

The next step in the program is to compute the divergent part of one
loop vertex with one boson and two fermions legs, compute the counterterms
needed and try to see whether they are sufficient to renormalize
all $n$-boson two fermion vertices.

\subsection{One loop contribution to two-fermion one-boson vertex}

In what follows we are going to compute the one loop contribution to
the one boson two fermions vertex following the same lines as in
Sec.\ref{parityeven}. The one-loop diagrams considered are the same
considered in \cite{aabe}, with the additional $F_{\mu\nu}$
structure.

All contributions to this vertex have been summarized in a
table contained in Appendix B. In this table we separate the different
structures in derivatives of the $U$ matrices, and we focus on their
$F$ structure.

The next step will be to use the relations
\begin{eqnarray*}
  \bar{x}_{\rho} ( A ) \theta ( A - B ) & = & - \int \bignone \mathd \tau
  \partial_{\tau} \theta ( A - \tau ) \bar{x}_{\rho} ( \tau ) \theta ( \tau -
  B )ŽÂŽÂŽº ,\\
  \bar{x}_{\rho} ( B ) \theta ( A - B ) & = & \int \bignone \mathd \tau \theta
  ( A - \tau ) \bar{x}_{\rho} ( \tau ) \partial_{\tau} \theta ( \tau - B )\ ,
\end{eqnarray*}
to convert the divergence in the one-boson vertex into a tree level
contribution in the Lagrangian.
The vertex operators extracted from this tree level expression are, of course,
directly related to the counterterms we are looking for and they read,
\begin{equation}
 \int \mathd \tau \frac{i}{2} \{ \bar{\psi}_L
   \bar{x}_{\rho} ( \tau ) \Phi_{\rho}^{( 1 )} \dot{\psi}_R -
   \dot{\bar{\psi}}_L \bar{x}_{\rho} ( \tau ) \Phi_{\rho}^{( 2 )} \psi_R \} \ ,
\end{equation}
where,
\begin{eqnarray*}
  \Phi_{\rho}^{( 1 )} & = & - ( \delta_{\eta \rho} - F_{\eta \rho} )
  \partial_{\eta} \delta U\\
  &  & - \Delta ( 0 ) \{ \frac{1}{4} \partial_{\eta} \partial_{\sigma} U
  \nonesep U^{\dag} \partial_{\lambda} U (( \delta_{\sigma \gamma} - F_{\sigma
  \gamma} ) ( \delta_{\eta \rho} + F_{\eta \rho} ) ( \delta_{\lambda \gamma} -
  F_{\lambda \gamma}) - [ F_{\eta \rho}, F_{\sigma \lambda}
  ](\delta_{\lambda\gamma}- F_{\lambda\gamma}))\\
  &  & \phantom{\Delta(0)}- \frac{1}{4} \partial_{\sigma} U \nonesep
  U^{\dag} \partial_{\lambda}
  \partial_{\eta} U ( ( \delta_{\sigma \gamma} + F_{\sigma \gamma} ) (
  \delta_{\eta \rho} - F_{\eta \rho} ) ( \delta_{\lambda \gamma} + F_{\lambda
  \gamma} ) -(\delta_{\sigma\gamma}+F_{\sigma\gamma}) [ F_{\lambda \gamma} \ ,
  F_{\eta \rho}] )\\
  &  & + \frac{1}{2} \partial_{\sigma} U \nonesep U^{\dag} \partial_{\eta} U
  \nonesep U^{\dag} \partial_{\lambda} U ( \delta_{\eta \rho} ( F_{\sigma
  \lambda} + F_{\lambda \sigma} ) - \delta_{\sigma \lambda} F_{\eta \rho} -
  F_{\sigma \gamma} F_{\eta \rho} F_{\lambda \gamma} + \frac{1}{2} [ F_{\eta
  \rho}, F_{\sigma \lambda} ] ( \delta_{\lambda \gamma} - F_{\lambda \gamma} )
  ) \}\\
  & \equiv & - ( \delta_{\eta \rho} - F_{\eta \rho} ) \partial_{\eta} \delta U -
  \phi_{\rho}  \ ,
\end{eqnarray*}
\begin{eqnarray*}
  \Phi_{\rho}^{( 2 )} & = & - ( \delta_{\eta \rho} + F_{\eta \rho} )
  \partial_{\eta} \delta U\\
  &  & + \Delta ( 0 ) \{ \frac{1}{4} \partial_{\eta} \partial_{\sigma} U
  \nonesep U^{\dag} \partial_{\lambda} U ( (\delta_{\sigma \gamma} - F_{\sigma
  \gamma} ) ( \delta_{\eta \rho} + F_{\eta \rho} ) ( \delta_{\lambda \gamma} -
  F_{\lambda \gamma} )- [ F_{\eta \rho}, F_{\sigma \gamma}
  ](\delta_{\lambda\gamma}- F_{\lambda\gamma}))\\
  &  &  \phantom{\Delta(0)}- \frac{1}{4} \partial_{\sigma} U
  \nonesep U^{\dag} \partial_{\lambda}
  \partial_{\eta} U ( ( \delta_{\sigma \gamma} + F_{\sigma \gamma} ) (
  \delta_{\eta \rho} - F_{\eta \rho} ) ( \delta_{\lambda \gamma} + F_{\lambda
  \gamma} ) -(\delta_{\sigma\gamma}+F_{\sigma\gamma}) [ F_{\lambda \gamma},
  F_{\eta \rho}] )\\
  &  & + \frac{1}{2} \partial_{\sigma} U \nonesep U^{\dag} \partial_{\eta} U
  \nonesep U^{\dag} \partial_{\lambda} U ( \delta_{\eta \rho} ( F_{\sigma
  \lambda} + F_{\lambda \sigma} ) - \delta_{\sigma \lambda} F_{\eta \rho} -
  F_{\sigma \gamma} F_{\eta \rho} F_{\lambda \gamma} + \frac{1}{2} [ F_{\eta
  \rho}, F_{\sigma \gamma} ] ( \delta_{\lambda \gamma} - F_{\lambda \gamma} )
  ) \}\\
  & = & - ( \delta_{\eta \rho} + F_{\eta \rho} ) \partial_{\eta} \delta U +
  \phi_{\rho} \ .
\end{eqnarray*}
Herein the following relation (induced from Eq.(\ref{algebra1})),
\begin{equation}
[F_{\eta\rho},F_{\sigma\gamma}](\delta_{\lambda\gamma}-F_{\lambda\gamma})
= (\delta_{\sigma\gamma}+F_{\sigma\gamma})[F_{\eta\rho},F_{\lambda\gamma}] \ ,
\end{equation}
has been used in order to make CP invariance manifest.

In these equations we have already separated two important parts.
The first part is a first variation of $U$ in the
Lagrangian's interaction part, so it is already under control. The
remainder of $\Phi_{\rho}^{( i )}$ is in fact the same in both $i=1$ and $2$
(only a sign makes a difference). Denoting the remainder as $\phi_\rho$ and
putting all together, one finds that the divergence is generated by the operators
\begin{eqnarray}
 && \int \mathd \tau [ \frac{i}{2} \{ \bar{\psi}_L
   \{\bar{x}_{\rho} ( \tau )\partial_{\rho} (-\delta
   U),\partial_\tau\}\psi_R + \frac{i}{2}\bar{\psi}_L
   \dot{\bar{x}}_{\rho} ( \tau )  F_{\eta\rho} \partial_{\eta}
   (-\delta U) \psi_R ] \\
&&\qquad\qquad\qquad+ \int \mathd \tau  \frac{i}{2}
    \bar{\psi}_L \dot{\bar{x}}_{\rho} ( \tau )\phi_{\rho}
   \psi_R \ .
\end{eqnarray}
The two first terms are already taken care of by the $\delta U$
counterterm, while the last one is dictating us the counterterms to
introduce to guarantee the finiteness of this vertex. Obviously, if 
we set $F_{\mu\nu}=0$ we recover the results of the spinless case.

\subsection{Counterterms}

In order to compensate the divergence in  $\phi_\rho$ we have to employ
the counterterm
\begin{eqnarray}
 &&\alpha' \int \mathd \tau \frac{i}{2} \bar{\psi}_L \dot{\bar{x}}_{\rho}
   \tilde{\phi}_{\rho} \psi_R \ , \label{resc}
\end{eqnarray}
where the rescaling on the dimensional constant $\alpha'$ has been
introduced to simplify some algebraic relations that follow. The structure
of $\tilde{\phi}_{\rho}$ is essentially determined by
$\phi_\rho$ and can be codified as follows
\begin{eqnarray}
  \tilde{\phi}_{\rho} & = &  \partial_{\sigma}\partial_{\eta} U
  \nonesep U^{\dag} \partial_{\lambda} U \nonesep A^{(1)}_{\sigma\eta
  \lambda \rho} + \partial_{\sigma} U \nonesep U^{\dag}
  \partial_{\eta}\partial_{\lambda}  U A^{(2)}_{\sigma \eta\lambda \rho}
  + \partial_{\sigma} U \nonesep U^{\dag} \partial_{\eta} U
  \nonesep U^{\dag} \partial_{\lambda} U \nonesep A^{(3)}_{\sigma \eta
  \lambda \rho} \ .
\label{counterterm1}
\end{eqnarray}
Evidently the operator coefficients  $A^{(1)}$ and $A^{(2)}$
are symmetric in a pair of indices,
\be
A^{(1)}_{\sigma\eta\lambda \rho} = A^{(1)}_{\eta\sigma\lambda \rho};\quad
 A^{(2)}_{\sigma \eta\lambda \rho} =  A^{(2)}_{\sigma\lambda\eta \rho},
\ee
being contracted with symmetric chiral field tensors.
One  can find the similarities of the counterterm \gl{counterterm1} with its
counterpart of Sec.\ref{parityeven} when $F_{\mu\nu}$ reduces to
$z\delta_{\mu\nu}$. In the present case the terms attached to each chiral
field $U$
structure are considerably more complex. The operator nature of $F_{\mu\nu}$
is a reason for a larger
set of coupling constants in the operator coefficients
$A^{(i)}_{\sigma\eta\lambda \rho}$'s.
As before the finite, renormalized part of this larger set of constants is to be
determined by the consistency equations of vanishing beta functions as well
as of local integrability of dimension-4 components of Eqs. of motion.
Let us present the actual form of $A^{(i)}_{\sigma\eta\lambda \rho}$ more explicitly
\begin{eqnarray}
  A^{(1)}_{\sigma\eta \lambda \rho} & = & A^{(1,r)}_{\sigma\eta\lambda \rho}
  - \frac{\Delta ( 0 )}{8\alpha'}\bigl(( \delta_{\sigma \gamma} - F_{\sigma
  \gamma} ) ( \delta_{\eta  \rho} + F_{\eta \rho} ) ( \delta_{\lambda
  \gamma} - F_{\lambda \gamma} )- [ F_{\eta \rho}, F_{\sigma \gamma}
  ](\delta_{\lambda\gamma}- F_{\lambda\gamma}) + \{\sigma
\leftrightarrow \eta\} \bigr)\nonumber\\
& = & \frac18 \Bigl( g^{(r)}
  - \frac{\Delta ( 0 )}{\alpha'}\Bigr)\bigl(( \delta_{\sigma \gamma} - F_{\sigma
  \gamma} ) ( \delta_{\eta  \rho} + F_{\eta \rho} ) ( \delta_{\lambda
  \gamma} - F_{\lambda \gamma} )- [ F_{\eta \rho}, F_{\sigma \gamma}
  ](\delta_{\lambda\gamma}- F_{\lambda\gamma}) + \{\sigma
\leftrightarrow \eta\} \bigr)\nonumber\\
  A^{(2)}_{\sigma \eta\lambda \rho} & = & A^{(2,r)}_{\sigma\eta \lambda\rho}
  + \frac{\Delta ( 0 )}{8\alpha'}\bigl( ( \delta_{\sigma \gamma} + F_{\sigma \gamma} ) (
  \delta_{\eta \rho} - F_{\eta \rho} ) ( \delta_{\lambda \gamma} + F_{\lambda
  \gamma} ) -(\delta_{\sigma\gamma}+F_{\sigma\gamma}) [ F_{\lambda \gamma},
  F_{\eta \rho}]+ \{\eta
\leftrightarrow \lambda\}\bigr)\nonumber\\
& = &- \frac18 \Bigl( \bar g^{(r)}
  - \frac{\Delta ( 0 )}{\alpha'}\Bigr)
 \bigl( ( \delta_{\sigma \gamma} + F_{\sigma \gamma} ) (
  \delta_{\eta \rho} - F_{\eta \rho} ) ( \delta_{\lambda \gamma} + F_{\lambda
  \gamma} ) -(\delta_{\sigma\gamma}+F_{\sigma\gamma}) [ F_{\lambda \gamma},
  F_{\eta \rho}]+ \{\eta
\leftrightarrow \lambda\}\bigr)\nonumber\\
  A^{(3)}_{\sigma \eta \lambda \rho} & = & A^{(3,r)}_{\sigma \eta \lambda \rho}
  - \frac{\Delta ( 0 )}{4\alpha'}\bigl( \delta_{\eta \rho} ( F_{\sigma \lambda}
  + F_{\lambda \sigma} ) - \delta_{\sigma \lambda} F_{\eta \rho} -
  F_{\sigma \gamma}   F_{\eta \rho} F_{\lambda \gamma} + \frac{1}{2} [
  F_{\eta \rho}, F_{\sigma\gamma} ] ( \delta_{\lambda \gamma} -
  F_{\lambda \gamma} ) \bigr)\nonumber\\
& = &\frac14 \Bigl( g^{(r)}_3
  - \frac{\Delta ( 0 )}{\alpha'}\Bigr) \bigl( \delta_{\eta \rho} ( F_{\sigma \lambda}
  + F_{\lambda \sigma} ) - \delta_{\sigma \lambda} F_{\eta \rho} -
  F_{\sigma \gamma}   F_{\eta \rho} F_{\lambda \gamma} + \frac{1}{2} [
  F_{\eta \rho}, F_{\sigma\gamma} ] ( \delta_{\lambda \gamma} -
  F_{\lambda \gamma} ) \bigr) \ ,
\label{counterterm2}
\end{eqnarray}
where  $A^{(i,r)}_{\sigma \eta \lambda \rho}$ are  renormalized
operators which depend on all finite parameters we referred above. As compared
to Sec.\ref{parityeven} these expressions contain the three similar constants
$ g^{(r)}, \bar g^{(r)}, g^{(r)}_3 $ but a
more complicated algebraic structure.

The actual composition of the $A^{(i,r)}_{\sigma\eta\lambda \rho}$ is
just a sum of products of the algebra elements $F_{\mu\nu}$
with independent finite
constants. We follow a minimal renormalization scheme and restrict
the form of the $A^{(i,r)}_{\sigma\eta\lambda\rho}$ by adopting only
the same $F$ combinations which appear in the
corresponding infinite part.

We notice also that CP invariance of the Lagrangian imposes the relations
\begin{eqnarray}
  A^{(1)}_{\sigma\eta\lambda\rho} =
-A^{(2)\dagger}_{\lambda\eta\sigma\rho}\ ,\qquad
A^{(3)}_{\sigma\eta\lambda\rho} =
-A^{(3)\dagger}_{\lambda\eta\sigma\rho} \ ,
\end{eqnarray}
While
the CP invariance of the divergent part holds manifestly due to the
Eq.(\ref{algebra1}), it is the CP invariance of the renormalized
part that we are interested in.
This condition applied to the parameterization \gl{counterterm2}
dictates that,
\begin{eqnarray}
  \bar g^{(r)} = -(g^{(r)})^* ,\qquad
g_3^{(r)} = (g_3^{(r)})^* ,\label{Herm1}
\end{eqnarray}
hence we end up with three real variables  Re $g^{(r)}$, Im $g^{(r)}$
and $g_3^{(r)}$ as in the scalar Lagrangian \gl{count}.

Now one must examine the two-boson two-fermion vertex in order to prepare
the two-loop renormalization of the fermion propagator.
We do not display this part of the
computation since it does not bring new counterterms and
the algebraic expressions
are rather cumbersome. All divergences in the one-loop two-boson
two-fermion vertex are proven to be  renormalized with the one-boson two-fermion
vertex counterterms. Thereby by translational invariance \cite{aabe} all
one-loop divergences in all n-boson two-fermion vertex are also
entirely renormalized. Thus the renormalization program at one loop is
completed. The inclusion of one-boson two-fermion counterterms
(\ref{counterterm1}), (\ref{counterterm2}) is sufficient to ensure the
complete renormalization at one loop.

\subsection{Dimension-4 divergences from one-loop counterterms
  and from two-loop contributions}

There are ten two-loop one-particle irreducible
diagrams which are listed
in \cite{aabe}.
The divergences in the propagator at two-loops can be
separated into five separate pieces
\begin{equation}
\theta(A-B)[ d_{I} + d_{II}+ d_{III}+ d_{IV}+ d_{V} ].
\end{equation}
The first and second piece contain the double pole divergence $ \Delta^2(0)$,
the third, fourth and fifth pieces contain the single pole divergence
$ \Delta (0)$.

The piece $d_I$ represents `the second variation', or
one-loop divergence in the one-loop divergence and it is removed by
the one-loop renormalization, hence it vanishes together with the
one-loop
$\beta$-function, {\it i.e.} when the equations of motion are imposed.

The second part represents the remaining terms of
order $\Delta^2(0)$ in two loop  diagrams
after subtraction of $d_I$.
This part is made of the contributions
generated by the one-loop
counterterm in the vertex with two fermions and one boson line,
after its insertion in a one-loop diagram.

$d_{III}$ contains those
single-pole divergences, proportional to
$\Delta (0)$, which are removed once the one-loop
renormalization of $U$ in the finite
nonlocal part of the fermion propagator at one loop
is taken into account.

The inclusion of the counterterms (\ref{counterterm1}),(\ref{counterterm2}) modifies in fact 
the fermion propagator adding terms of higher
order in derivatives (of dimension 4 in the count of Chiral
Perturbation Theory). Eventually the following divergent contributions to the
propagator $\sim \Delta(0)$ arise from the finite, renormalized
part (\ref{counterterm2}) of the counterterms \gl{counterterm1} when introduced 
in one-loop diagrams
\begin{center}
  \begin{tabular}{|c|c|c|}
    \hline
    Coefficient & U structure & F structure\\
    \hline
    $+ \frac{1}{8} \alpha'\theta ( A - B ) \Delta ( 0 )$ & $U^{\dag}
    \partial_{\alpha} \partial_{\sigma} U \nonesep U^{\dag} \partial_{\lambda}
    \partial_{\beta} U \nonesep U^{\dag}$ & 0\\
    \hline
    $+ \frac{1}{8}\alpha' \theta ( A - B ) \Delta ( 0 )$ & $U^{\dag}
    \partial_{\alpha} \partial_{\sigma} U \nonesep U^{\dag} \partial_{\lambda}
    U \nonesep U^{\dag} \partial_{\beta} U \nonesep U^{\dag}$ & $- 2
    A^{(1,r)}_{ \sigma\alpha \lambda \mu} ( \delta_{\beta \mu} - F_{\beta \mu} )$\\
    \hline
    $+ \frac{1}{8}\alpha' \theta ( A - B ) \Delta ( 0 )$ & $U^{\dag}
    \partial_{\alpha} U \nonesep U^{\dag} \partial_{\sigma} U \nonesep
    U^{\dag} \partial_{\lambda} \partial_{\beta} U \nonesep U^{\dag}$ & $+ 2 (
    \delta_{\alpha \mu} + F_{\alpha \mu} ) A^{(2,r)}_{\sigma\beta \lambda\mu}$\\
    \hline
    $+ \frac{1}{8} \alpha'\theta ( A - B ) \Delta ( 0 )$ & $U^{\dag}
    \partial_{\alpha} U \nonesep U^{\dag} \partial_{\sigma} \partial_{\beta}
    U \nonesep U^{\dag} \partial_{\lambda} U \nonesep U^{\dag}$ & $2 ( (
    \delta_{\alpha \mu} + F_{\alpha \mu} ) A^{(1,r)}_{\sigma \beta\lambda\mu} -
    A^{(2,r)}_{\alpha \sigma \beta \mu} ( \delta_{\lambda \mu} - F_{\lambda \mu} )
    )$\\
    \hline
    $+ \frac{1}{8}\alpha' \theta ( A - B ) \Delta ( 0 )$ & $U^{\dag}
    \partial_{\alpha} U \nonesep U^{\dag} \partial_{\sigma} U \nonesep
    U^{\dag} \partial_{\beta} U \nonesep U^{\dag} \partial_{\lambda} U
    \nonesep U^{\dag}$ & $2 ( ( \delta_{\alpha \mu} + F_{\alpha \mu} )
    A^{(3,r)}_{\sigma \beta \lambda \mu} - A^{(3,r)}_
{\alpha \sigma \beta \mu} (
    \delta_{\lambda \mu} - F_{\lambda \mu} ) )$\\
    \hline
  \end{tabular}\\
\medskip

{\bf Table 1}. Divergent contribution to the propagator from the vertex counterterms.
\medskip

\end{center}
In  $d_{IV}$ we include the  divergences that are eliminated when
the additional counterterms in the one-boson vertices (those
proportional to $A^{(i)}$) are included
in the finite part of the one-loop fermion propagator.
One can check (in a way similar to \cite{aabe}) that all
terms in the two loop fermion propagator linear
in $\Delta(0)$ and in  $\Delta (A,B)$ belong either
to  $d_{III}$ or to $d_{IV}$.

Finally, some single-pole divergences remain and they are gathered in
$d_{V}$.
Namely, there are
 divergences linear in
 $\Delta(0)$ which come from the double integral in irreducible two-loop
diagrams with maximal number of vertices (overlapping divergences),
\begin{equation}
J(A-B)=  \int^A_B d\tau_1 \int^{\tau_1}_B d\tau_2
\partial_{\tau_1}\Delta (\tau_1-\tau_2) \partial_{\tau_2}\Delta
(\tau_1- \tau_2)  = 2\alpha' \Delta (0) + \mbox{\rm finite part}. \label{Jab}
\end{equation}
These operators are described in the table

\begin{center}
\begin{tabular}{|c|c|c|}
    \hline
    Coefficient & U structure & F structure\\
    \hline
    $+\frac{1}{8} \alpha' \theta ( A - B ) \Delta (0)$ & $U^{\dag} \partial_{\alpha}
    \partial_{\sigma} U \nonesep U^{\dag} \partial_{\lambda} \partial_{\beta}
    U \nonesep U^{\dag}$ & $\bigl( F_{\sigma \lambda} F_{\beta \alpha} -
    \delta_{\alpha \beta} F_{\sigma \gamma} F_{\lambda \gamma} +
    \{\alpha\leftrightarrow \sigma, \lambda\leftrightarrow\beta\}\bigr) $\\
    \hline
    $ +\frac{1}{8} \alpha' \theta ( A - B ) \Delta (0)$ & $U^{\dag} \partial_{\alpha}
    \partial_{\sigma} U \nonesep U^{\dag} \partial_{\lambda} U \nonesep
    U^{\dag} \partial_{\beta} U \nonesep U^{\dag}$ & $\bigl( F_{\alpha \rho} (
    \delta_{\sigma \lambda} - F_{\lambda \sigma} ) ( \delta_{\beta \rho} +
    F_{\beta \rho} )$\\
    &  & $- F_{\sigma \gamma} ( \delta_{\lambda \gamma} - F_{\lambda \gamma}
    ) ( \delta_{\alpha \beta} + F_{\beta \alpha} ) + \{\alpha\leftrightarrow
    \sigma \} \bigr)$\\
    \hline
    $ +\frac{1}{8} \alpha' \theta ( A - B )  \Delta (0)$ & $U^{\dag} \partial_{\alpha}
    U \nonesep U^{\dag} \partial_{\sigma} U \nonesep U^{\dag}
    \partial_{\lambda} \partial_{\beta} U \nonesep U^{\dag}$ & $\bigl( (
    \delta_{\alpha \beta} - F_{\alpha \beta} ) ( \delta_{\sigma \gamma} +
    F_{\sigma \gamma} ) F_{\lambda \gamma}$\\
    &  & $- ( \delta_{\alpha \rho} - F_{\alpha \rho} ) ( \delta_{\sigma
    \lambda} + F_{\sigma \lambda} ) F_{\beta \rho} +
\{\lambda\leftrightarrow\beta\} \bigr)$\\
    \hline
    $ +\frac{1}{8} \alpha' \theta ( A - B ) \Delta (0)$ & $U^{\dag}
    \partial_{\alpha} U \nonesep U^{\dag} \partial_{\sigma} U \nonesep
    U^{\dag} \partial_{\lambda} U \nonesep U^{\dag} \partial_{\beta} U
    \nonesep U^{\dag}$ & $\bigl( (\delta_{\alpha \rho} + F_{\alpha \rho} ) (
    \delta_{\sigma \gamma} + F_{\sigma \gamma} ) ( \delta_{\lambda \gamma} -
    F_{\lambda \gamma} )$\\
    &  & $\times ( \delta_{\beta \rho} - F_{\beta \rho} ) - ( \delta_{\alpha \rho} -
    F_{\alpha \rho} )$\\
    &  & $\times( \delta_{\sigma \gamma} + F_{\sigma \gamma} ) ( \delta_{\lambda
    \rho} - F_{\lambda \rho} ) ( \delta_{\beta \gamma} +
F_{\beta \gamma} )\bigr)$\\
    \hline
  \end{tabular}\\
\medskip

{\bf Table 2}. Relevant part of arising overlapping divergences.
\medskip

\end{center}
The terms in $d_{V}$ survive
after adding all the counterterms and together with Table 1 
are the only new genuine divergences that can
contribute to the beta function (single poles).
It must therefore be added to
the equation of motion at the next order in the $\alpha^\prime$ expansion.
Thus two sets of operators listed in Tables 1 and 2 form the genuine
contribution $U^\dagger\delta^{(4)} U U^\dagger$ of chiral dimension 4 into
the beta-functional of the chiral field renormalization. To this order the
condition of conformal invariance reads
\be
\delta^{(2)} U+ \delta^{(4)} U = 0,
\ee
and this equation must be identified with an equation of local chiral
dynamics if we deal with the Goldstone boson physics of pseudoscalar mesons.
However such an identification is not unique as one may have certain terms
in  $\delta^{(4)} U$ vanishing on the mass-shell
$\delta^{(2)} U = 0$ . This is a logic of  Chiral Perturbation Theory.
Therefore in the comparison of the beta-functional   $\delta^{(4)} U$ and a relevant
functional of local chiral dynamics one must include all possible
operators vanishing on-shell.

\section{Local integrability of dimension-4 part of Eqs. of motion}

If the corresponding terms with four derivatives that we have  found
in the previous section
 originate from a dimension-four operators in a quasi-local
effective Lagrangian then certain constraints are to be imposed on
the constants $A^{(i,r)}_{\mu}$.

On mass-shell such a Lagrangian has only three terms compatible with
the chiral symmetry if we employ the dimension-two equations of motion
(\ref{resym}),
\begin{eqnarray}
S^{(4)} &=& \frac {f_\pi^2 \alpha'}{8}  \int d^4 x \,
\mbox{tr}\Biggl( K_1 \partial_\mu U \partial_\rho  U^\dagger 
\partial_{\mu} U \partial_{\rho} U^\dagger
+ K_2  \partial_\mu U \partial_{\mu} U^\dagger \partial_{\rho} U
\partial_{\rho} U^\dagger\nonumber\\
&& - \frac15 \int^1_0 dx_5  K_3  \epsilon_{ABCDE} \partial_A U  U^\dagger \partial_B U
U^\dagger   \partial_C U  U^\dagger  \partial_{D} U   U^\dagger  \partial_E U  U^\dagger
\Biggr), \label{chirdim4}
\end{eqnarray}
the last operator being the celebrated Wess-Zumino-Witten term \cite{witt}. The
capital Latin indices\\ $A,B,C,D,E = 1,\ldots 5$ mark tensors in the five-dimensional space with a
compact fifth coordinate $0 \leq x_5 \leq 1$ whereas the Greek indices
mark the four dimensional Euclidean coordinates. The fully
antisymmetric tensor  $\epsilon_{ABCDE}$ is conventionally normalized
to  $\epsilon_{12345} = 1$. It is assumed also that
$U(x_5=0) = 1$ and $U(x_5=1) \equiv U(x_\mu)$. The normalization in the front
of the integral of Eq.\gl{chirdim4} is chosen to simplify the forthcoming
consistency conditions.

The terms
$$\partial_\mu^2 U \partial_\rho  U^\dagger \partial_{\rho} U U^\dagger,\qquad
\partial_\mu^2 U \partial_\rho^2  U^\dagger,\qquad  (\partial_\mu^2)^2 U   U^\dagger,\qquad
\partial_\mu\partial_\rho U \partial_{\mu} \partial_{\rho} U^\dagger $$
which are in principle acceptable are reduced to the set (\ref{chirdim4})
with the help
of integration by parts in the
action and  of  the dimension-two equations of motion (\ref{resym}) (on-shell
conditions).

Variation of the previous Lagrangian gives the following
addition to the equations of motion,
\begin{eqnarray}
\frac{1}{f_\pi^2}\frac{\delta S^{(4)}}{\delta U} &=& -
\frac{\alpha'}{8}  U^\dagger \left\{4 K_1\left[
\partial_\mu\partial_\rho U  U^\dagger \partial_{\mu} U  U^\dagger \partial_{\rho} U
+  \partial_\mu U  U^\dagger\partial_\rho U  U^\dagger \partial_{\mu}
\partial_{\rho} U\right.\right.\nonumber\\
&&\left.\left.
-   \partial_\mu U U^\dagger\partial_\rho U  U^\dagger \partial_{\mu} U U^\dagger
\partial_{\rho} U
-2 \partial_\mu U  U^\dagger \partial_{\rho} U  U^\dagger \partial_{\rho} U U^\dagger
\partial_\mu U\right.\right.\nonumber\\
&&\left.\left.
+ \partial_\mu U U^\dagger\partial_\rho^2 U  U^\dagger \partial_{\mu} U \right]
\right.\nonumber\\
&&\left. + 2 K_2\left[\partial_\mu\partial_\rho U  U^\dagger \partial_{\mu} U
U^\dagger \partial_{\rho} U
+  \partial_\mu U  U^\dagger\partial_\rho U  U^\dagger \partial_{\mu}
\partial_{\rho} U \right.\right.\nonumber\\
&&\left.\left. +
2 \partial_{\mu} U U^\dagger \partial_\mu\partial_\rho U  U^\dagger
\partial_{\rho} + \partial_\mu^2 U U^\dagger \partial_{\rho} U U^\dagger
\partial_{\rho} U
+ \partial_\mu U U^\dagger \partial_{\mu} U^\dagger \partial_{\rho}^2 U
\right.\right.\nonumber\\
&&\left.\left.
-  2 \partial_\mu U U^\dagger\partial_\rho U  U^\dagger
\partial_{\mu} U U^\dagger\partial_{\rho} U
- \partial_\mu U  U^\dagger \partial_{\rho} U  U^\dagger \partial_{\rho} U U^\dagger
\partial_\mu U\right.\right.\nonumber\\
&&\left.\left.
-3 \partial_\mu U U^\dagger\partial_{\mu} U U^\dagger \partial_{\rho} U U^\dagger
\partial_{\rho} U\right] + K_3 \epsilon_{\alpha \sigma \lambda \beta}
\partial_\alpha U U^\dagger\partial_{\sigma} U U^\dagger
\partial_{\lambda} U U^\dagger
\partial_{\beta} U
\right\}U^\dagger. \label{dimen4}
\end{eqnarray}
Now we proceed to the comparison of the beta-functional  $\delta^{(4)} U$ and
the four-derivative part of Eqs. of
chiral dynamics \gl{dimen4}.  As it has been elucidated in the previous section
one expects the entire identification on the dimension 2 mass-shell
({i.e.} after applying of the $O(p^2)$ Eqs. of motion). Off-shell one must extend \gl{dimen4} to
the following general set of operators and coefficients
for the various chiral
field structures,
\newpage

\begin{center}
  \begin{tabular}{|c|c|}
    \hline
    U structure & F structure\\
    \hline
   $\frac{\alpha'}{8} U^{\dag} \partial_{\alpha} \partial_{\sigma} U \nonesep U^{\dag}
    \partial_{\lambda} \partial_{\beta} U \nonesep U^{\dag}$ &
    $C_0\delta_{\alpha\sigma} \delta_{\beta\lambda} +
    B^{(0)}_{\alpha\sigma}\delta_{\beta\lambda}
    +\delta_{\alpha\sigma} B^{(1)}_{\beta\lambda}$ \\
    \hline
  $\frac{\alpha'}{8}  U^{\dag} \partial_{\alpha} \partial_{\sigma} U \nonesep U^{\dag}
    \partial_{\lambda} U \nonesep U^{\dag} \partial_{\beta} U \nonesep
    U^{\dag}$ & $- ( 2 K_1 + K_2 ) ( \delta_{\alpha \beta} \delta_{\sigma
    \lambda} + \delta_{\sigma \beta} \delta_{\alpha \lambda} )
    - B^{(0)}_{\alpha\sigma} \delta_{\beta\lambda}$\\
    &$+\delta_{\alpha\sigma}(C_1\delta_{\beta\lambda}+ B^{(2)}_{\beta\lambda})$\\
    \hline
 $ \frac{\alpha'}{8}  U^{\dag} \partial_{\alpha} U \nonesep U^{\dag} \partial_{\sigma} U
    \nonesep U^{\dag} \partial_{\lambda} \partial_{\beta} U \nonesep U^{\dag}$
    & $- ( 2 K_1 + K_2 ) ( \delta_{\alpha \beta} \delta_{\sigma \lambda} +
    \delta_{\alpha \lambda} \delta_{\sigma \beta} )
    - \delta_{\alpha\sigma} B^{(1)}_{\beta\lambda} $\\
    &$+\bigl(C_2\delta_{\alpha\sigma}+ B^{(3)}_{\alpha\sigma}\bigr)\delta_{\beta\lambda}$\\
    \hline
   $\frac{\alpha'}{8} U^{\dag} \partial_{\alpha} U \nonesep U^{\dag} \partial_{\sigma}
    \partial_{\lambda} U \nonesep U^{\dag} \partial_{\beta} U \nonesep
    U^{\dag}$ & $- 2 K_2 ( \delta_{\alpha \sigma} \delta_{\beta \lambda} +
    \delta_{\alpha \lambda} \delta_{\beta \sigma} )
    +\delta_{\sigma\lambda}(C_3\delta_{\alpha\beta}
    +B^{(4)}_{\alpha\beta})$\\
    \hline
  $\frac{\alpha'}{8}  U^{\dag} \partial_{\alpha} U \nonesep U^{\dag} \partial_{\sigma} U
    \nonesep U^{\dag} \partial_{\lambda} U \nonesep U^{\dag} \partial_{\beta}
    U \nonesep U^{\dag}$ & $ (2 ( 2 K_1 + K_2 )-C_3) \delta_{\alpha \beta}
    \delta_{\sigma \lambda} + (2 K_2-C_0-C_1 -C_2) \delta_{\alpha \sigma}
    \delta_{\beta\lambda}$\\
 &$+ 4 ( K_1 + K_2 ) \delta_{\alpha \lambda}
    \delta_{\sigma \beta}
   + K_3\epsilon_{\alpha\sigma\lambda\beta}
-\delta_{\alpha\sigma}B^{(2)}_{\beta\lambda}
   $\\
    &$ -B^{(3)}_{\alpha\sigma} \delta_{\beta\lambda}
    -\delta_{\sigma\lambda} B^{(4)}_{\alpha\beta}$\\
    \hline
  \end{tabular}\\
\medskip

{\bf Table 3}. Dimension 4 operators from a local chiral Lagrangian
supplemented by the 'additional' off-shell contributions
. See the text for an explanation of their necessity .
\medskip

\end{center}
In this Table the numeric, $C_i$ and operator, $B_i$ coefficients have been
inserted so that they compensate each other in the total sum on the
mass-shell. Evidently the two operators $ B^{(0)}, B^{(1)}$ are symmetric
tensors,
\be
 B^{(0)}_{\alpha\beta} = B^{(0)}_{\beta\alpha}\ ,\quad
B^{(1)}_{\alpha\beta} = B^{(1)}_{\beta\alpha}\ .
\ee

Let us identify this parameterization of local Lagrangian descendants
with the coupling
constants and operator coefficients
arising from the vertices in Tables 1 and 2. In particular
the coefficients $C_i$ and $B_i$
admit  weaker algebraic restrictions on
the operators $F_{\mu\nu}$. The pertinent
consistency equations read, in the same order as in the preceding table,

\begin{eqnarray}
 ( F_{\sigma \lambda} F_{\beta \alpha} -
  \delta_{\alpha \beta}
  F_{\sigma \gamma} F_{\lambda \gamma} ) +  (F_{\alpha \lambda}
  F_{\beta \sigma} - \delta_{\sigma \beta} F_{\alpha \gamma} F_{\lambda
  \gamma} ) \text{} &  & \nonumber\\
  +  ( F_{\sigma \beta} F_{\lambda \alpha} -
  \delta_{\alpha \lambda}
  F_{\sigma \gamma} F_{\beta \gamma} ) +  ( F_{\alpha\beta}
  F_{\lambda \sigma} - \delta_{\sigma \lambda} F_{\alpha \gamma}
  F_{\beta
  \gamma} ) & = &C_0\delta_{\alpha\sigma} \delta_{\beta\lambda} +
    B^{(0)}_{\alpha\sigma}\delta_{\beta\lambda}
    +\delta_{\alpha\sigma} B^{(1)}_{\beta\lambda}\ ;
  \label{alg1}
\end{eqnarray}

\begin{eqnarray}
   ( F_{\alpha \rho} ( \delta_{\sigma \lambda} - F_{\lambda \sigma} ) (
  \delta_{\beta \rho} + F_{\beta \rho} ) - F_{\sigma \gamma} ( \delta_{\lambda
  \gamma} - F_{\lambda \gamma} ) ( \delta_{\alpha \beta} + F_{\beta \alpha} )
  ) &  & \nonumber\\
  +  ( F_{\sigma \rho} ( \delta_{\alpha \lambda} - F_{\lambda \alpha} )
  ( \delta_{\beta \rho} + F_{\beta \rho} ) - F_{\alpha \gamma} (
  \delta_{\lambda \gamma} - F_{\lambda \gamma} ) ( \delta_{\sigma \beta} +
  F_{\beta \sigma} ) ) &  & \label{alg2}\\
- 2 A^{(1,r)}_{ \sigma\alpha \lambda \mu}
( \delta_{\beta \mu} - F_{\beta \mu} )
 & = & - ( 2 K_1 + K_2 ) ( \delta_{\alpha \beta} \delta_{\sigma
    \lambda} + \delta_{\sigma \beta} \delta_{\alpha \lambda} )\nonumber\\
    &&- B^{(0)}_{\alpha\sigma} \delta_{\beta\lambda}
    +\delta_{\alpha\sigma}(C_1\delta_{\beta\lambda}+ B^{(2)}_{\beta\lambda})\ ;
  \nonumber
\end{eqnarray}

\begin{eqnarray}
   ( ( \delta_{\alpha \beta} - F_{\alpha \beta} ) ( \delta_{\sigma
  \gamma} + F_{\sigma \gamma} ) F_{\lambda \gamma} - ( \delta_{\alpha \rho} -
  F_{\alpha \rho} ) ( \delta_{\sigma \lambda} + F_{\sigma \lambda} ) F_{\beta
  \rho} ) &  & \nonumber\\
  +  ( ( \delta_{\alpha \lambda} - F_{\alpha \lambda} ) (
  \delta_{\sigma \gamma} + F_{\sigma \gamma} ) F_{\beta \gamma} - (
  \delta_{\alpha \rho} - F_{\alpha \rho} ) ( \delta_{\sigma \beta} + F_{\sigma
  \beta} ) F_{\lambda \rho} ) &  &\\
+ 2 ( \delta_{\alpha \mu} + F_{\alpha \mu} )
A^{(2,r)}_{\sigma\beta \lambda\mu} & = & - ( 2 K_1 + K_2 ) ( \delta_{\alpha \beta} \delta_{\sigma \lambda} +
    \delta_{\alpha \lambda} \delta_{\sigma \beta} ) \nonumber\\
   && - \delta_{\alpha\sigma} B^{(1)}_{\beta\lambda}
    +\bigl(C_2\delta_{\alpha\sigma}+ B^{(3)}_{\alpha\sigma}\bigr)\delta_{\beta\lambda}\ ;
  \nonumber \label{alg3}
\end{eqnarray}

\begin{eqnarray}
 2 ( (\delta_{\alpha \mu} + F_{\alpha \mu} )
A^{(1,r)}_{\sigma \beta\lambda\mu} -
    A^{(2,r)}_{\alpha \sigma \beta \mu} ( \delta_{\lambda \mu} -
F_{\lambda \mu} )) & =
  & - 2 K_2 ( \delta_{\alpha \sigma} \delta_{\beta \lambda} +
    \delta_{\alpha \lambda} \delta_{\beta \sigma} ) \label{alg4} \\
   && +\delta_{\sigma\lambda}(C_3\delta_{\alpha\beta}
    +B^{(4)}_{\alpha\beta})\ ;
  \nonumber
\end{eqnarray}

\begin{eqnarray}
  ( \delta_{\alpha \rho} + F_{\alpha \rho} ) ( \delta_{\sigma \gamma} +
  F_{\sigma \gamma} ) ( \delta_{\beta \gamma} - F_{\beta \gamma} ) (
  \delta_{\lambda \rho} - F_{\lambda \rho} ) &  & \nonumber\\
  + ( \delta_{\alpha \rho} - F_{\alpha \rho} ) ( \delta_{\sigma \gamma} +
  F_{\sigma \gamma} ) ( \delta_{\beta \rho} - F_{\beta \rho} ) (
  \delta_{\lambda \gamma} + F_{\lambda \gamma} ) &  & \nonumber\\
+  2 ( ( \delta_{\alpha \mu} + F_{\alpha \mu} )
    A^{(3,r)}_{\sigma \beta \lambda \mu} - A^{(3,r)}_{\alpha \sigma \beta \mu}
( \delta_{\lambda \mu} - F_{\lambda \mu} ) ) & = & (2 ( 2 K_1 + K_2 )-C_3) \delta_{\alpha \beta}
    \delta_{\sigma \lambda} \\&&
+ (2 K_2-C_0-C_1 -C_2) \delta_{\alpha \sigma}
    \delta_{\beta\lambda}\nonumber\\
 &&+ 4 ( K_1 + K_2 ) \delta_{\alpha \lambda}
    \delta_{\sigma \beta}
   + K_3\epsilon_{\alpha\sigma\lambda\beta}\nonumber\\
&&-\delta_{\alpha\sigma}B^{(2)}_{\beta\lambda}
    -B^{(3)}_{\alpha\sigma} \delta_{\beta\lambda}
    -\delta_{\sigma\lambda} B^{(4)}_{\alpha\beta}\ . \nonumber
  \label{alg5}
\end{eqnarray}

These equations dictate the consistency conditions for the algebra of 
the operators $F_{\mu\nu}$ and
bound the values of the low-energy constants  $K_1,
K_2, K_3$ in the chiral Lagrangian. The simplest solution
for $F_{\mu\nu} = z\delta_{\mu\nu}$ of this set of conditions was
obtained in \cite{aabe} and briefly described in Section 2. This solution, however,
does not describe the spin degrees of freedom as implicitly assumes that quarks are scalar
objects under rotations.
The next simplest hypothesis
is that $F_{\mu\nu}$ has an antisymmetric part proportional to
$\sigma_{\mu\nu}$. Indeed,  the coupling $\bar\psi_L\sigma_{\mu\nu}\psi_R  \times
\dot x_\mu\partial_\nu U$ intuitively reflects the coupling between the 
string angular momentum (at the boundary) assuming that 'quarks' have $s=1/2$ and
the angular momentum of the $U$-field. 
As we have seen at the beginning of section 5 this
not compatible with the one-loop renormalization properties of the model.
We are then forced to conclude that the Grassmann variables are not in a state of well defined
spin $s=1/2$.

It is possible to use the previous set of equations \gl{alg1}--\gl{alg5} to further constrain the
operators $F_{\mu\nu}$. This is a rather non-trivial task.
In next Section we explore some identities for  operators $F_{\mu\nu}$ and
discuss possible realizations of this algebra. Even though we do not have
a final answer and, in a sense, Eqs.\gl{algebra1}, \gl{alg1}--\gl{alg5} are our final result, 
the problem is
interesting enough, deserving more detailed considerations.

\section{Algebra considerations}

It is probably worth to recapitulate where we stand.

The elimination of all divergences at the one-loop order requires,
in addition to redefining the unitary matrix $U$, additional counterterms
that are given in Eqs.\gl{counterterm1},  \gl{counterterm2} in terms of a certain number of constants
$g^{(r)}, \bar g^{(r)}g_3^{(r)}$.
In spite of the rather large number of structures involving  $F_{\mu\nu}$,
only three independent combinations
 appear in the counterterms described in Eqs.\gl{counterterm1},  \gl{counterterm2}. 
This is somewhat reminiscent of
the situation without the spin structures, where three additional constants
$g_{i,r} $, each one accompanying a different chiral structure, are engaged.
The complication here lies of course in the fact that the  $F_{\mu\nu}$ are operators
taking values in some algebra yet to be specified.

Renormalizing the two-loop order propagator (i.e. $U$) needs taking
all these one-loop counterterms. Adding all the contributions up leads to the
conditions listed in Tables 1 -- 3  and to the set of Eqs. \gl{alg1}-- \gl{alg5}.

As explained,
these relations equate the single pole divergent part of the fermion
propagator (a combination of chiral fields $U$, their derivatives, and operators
 $F_{\mu\nu}$)  with the equivalent terms arising in equations of motion
derived from the local Chiral Lagrangian (\ref{chirdim4}). These equations of motion
involve chiral fields and their derivatives, but not $F_{\mu\nu}$. If we insist, as we should, in
making the two set of expressions equivalent this naturally brings about
new relations involving the $F_{\mu\nu}$.

Through these equations we can learn more about the form of the $F_{\mu\nu}$
operator matrix and thus the way the spinor interaction degrees of freedom are
implemented into this $F_{\mu\nu}$ operator, and of course, when possible,
fix as much as we can the value of $K_1$, $K_2$ and $K_3$.
These relations stem from the requirements of chiral invariance and locality of the
effective action and they should be understood as restrictions that these conditions
place on the algebra that the $F_{\mu\nu}$ satisfy.

To be specific, from Subsec. 5.2 , Eq. \gl{algebra1} we have
\begin{equation}
 F_{\sigma \gamma} F_{\lambda \gamma}= -\delta_{\sigma \lambda} +
 F_{\sigma \lambda} - F_{\lambda \sigma};\quad  F_{\lambda\gamma}
F_{\lambda\gamma}= - 4\ .\label{Eq_oneloop_1}
\end{equation}
Next the fulfillment of Eq. \gl{alg1} turns out to be very crucial as it
removes the chiral field structure which is a serious obstruction for local
integrability of Eqs. of motion. Therefrom,
after contracting two of the indices with $\delta_{\alpha\beta}$, we obtain
\begin{equation}
F_{\gamma\sigma}F_{\gamma\lambda}=-(9+C_0)\delta_{\sigma\lambda}
-5 (F_{\sigma\lambda} -F_{\lambda\sigma}) - \tr{F} F_{\sigma\lambda}
-F_{\lambda\sigma} \tr{F}
+ B^{(0)}_{\sigma\lambda}
    + B^{(1)}_{\sigma\lambda}\  .
\label{contr1}
\end{equation}
As the components of the operator $F_{\sigma\lambda}$ are antihermitian it
comes out from \gl{contr1} that,
\be
C_0 = (C_0)^*;\quad \bigl( B^{(0)}_{\sigma\lambda}
    + B^{(1)}_{\sigma\lambda}\bigr)^\dagger =  B^{(0)}_{\sigma\lambda}
    + B^{(1)}_{\sigma\lambda}\ .
\ee
As well from the further contraction of indices $\sigma =\lambda$ one
determines the trace of the operator $F_{\sigma\lambda}$,
\be
(\tr{F})^2 = 2 C_0 - 16 + \tr{ B^{(0)} + B^{(1)} } .
\ee
We however stress that in general it represents an operator relation when one
of the traces is not a c-number.

Finally, a non-equivalent contraction  allows us to fix the symmetric part of
twist-contracted products of $F_{\gamma\sigma}$,
\begin{equation}
F_{\gamma\sigma}F_{\lambda\gamma}+F_{\gamma\lambda}F_{\sigma\gamma}=
2(C_0 - 1 + \frac14 \tr{ B^{(0)}}  )\delta_{\sigma\lambda}
 + 2 B^{(1)}_{\sigma\lambda}\ ,\label{contr2}
\end{equation}
that allows for the determination of twisted normalization of the operator
$F_{\sigma\lambda}$,
\be
F_{\gamma\lambda}F_{\lambda\gamma} = 4 (C_0 - 1) + \tr{ B^{(0)}+  B^{(1)}} =
28 + 2  (\tr{F})^2\ .
\ee
All these algebraic relations originate from the requirement
of local integrability
of the would-be equations of motion. Notice that the last one \gl{contr2} does not
give us an explicit algebraic expression for the antisymmetric
part. An ansatz admitting  lineal in $F$ right-hand  parts of
Eqs. \gl{contr1}, \gl{contr2}  would
close the algebra. However it
happens to lead to a definite contradiction when the associativity of the
algebra of contracted and twist-contracted products of three $F_{\gamma\sigma}$ is examined. Hence
the ansatz is not correct and the algebra does not close.

Unfortunately, at the end of the day, we shall not have an explicit realization of the $F_{\mu\nu}$
satisfying all the previous requirements.
Some obvious possibilities are however ruled out. We have
already mentioned that the attempt of identifying the antisymmetric part of $F_{\mu\nu}$
with $\sigma_{\mu\nu}$ fails (see Subsection 5.1).
It is somewhat more surprising that if
Eqs. \gl{algebra1}, \gl{alg1} -- \gl{alg5}  are to be imposed, 
the described algebra spanned by the $F_{\mu\nu}$ does not
close, so it must necessarily be embedded in a larger algebra.

We are then forced to somewhat loosen the requirement of closure of the
algebra.
At this point, we discontinue the analysis of the implications of the
algebraic relations  \gl{algebra1}, \gl{contr1}, \gl{contr2}.  
We regard these equations as constraints that the algebra
of the $F$'s must satisfy in order to provide consistent propagation of the hadronic string
in a chirally non-invariant vacuum when the spin degrees of freedom are taken into account.
The remaining relations  \gl{alg2} -- \gl{alg5} are rather tools for the
estimation of all coupling constants introduced on the boundary as well as the
chiral constants $K_1, K_2, K_3$. This program nevertheless requires first
to discover the
algebra
of $F_{\mu\nu}$ to be predictive .

\section{Conclusions}

In this work we have analyzed in detail the conditions that the effective string conceived to
describe the interactions between quarks at long distances in QCD must meet. An essential ingredient
for this string is the assumption that in the real QCD vacuum chiral symmetry is broken and
the propagation takes place in a background of $\Pi$-on fields (not states on the
Regge trajectories). The condition of locality, chiral symmetry and conformal invariance place strong
constraints on this background, eventually leading to vanishing beta functionals to be interpreted
as equations of motion of the non-linear sigma model describing $\Pi$-on interactions.

The work reported here dwells on a previous analysis where quarks (represented by Grassmann
variables living on a line) were consider to be scalars. But spin is indeed an important variable
in Regge analysis (let us recall here the existence of the so-called $S$ and $D$ Regge trajectories).
More importantly, it is not difficult to see that without considering angular momentum, the
odd (internal) parity of the $\Pi$-on Lagrangian (i.e. the Wess-Zumino-Witten action) will never
be obtained as one of the byproducts of requiring conformal invariance.

In the preceding pages a number of new results have been obtained. We have managed to couple
an external gauge field and in this way to derive the covariant $O(p^2)$ equations of motion.
The analysis of the Wess-Zumino-Witten action in dimension 2 turns out to be rather straightforward
and it reproduces well the expected results.

Angular momentum in two dimensions is somewhat special and this is reflected in its realization
in terms of gamma matrices. In fact the calculation can be fully reformulated
using  scalar variables.
When proceeding to the four-dimensional case, things become rather more involved. We construct the
general coupling that involves some operator coupling $F_{\mu\nu}$ 
(acting on the angular momentum degrees
of freedom of the quarks). Consistency conditions of the string propagation
indeed remarkably seem to suggest
that the quarks are not in a definite state of angular momentum.
A deeper reason may be in that hadron string realizes the Reggeization
of meson states which, in the spirit of quark-hadron duality,
presumably follows from a Reggeization of quarks and
gluons as it happens in the semi-hard high-energy scattering in QCD
\cite{Lip}. If such a quark-hadron duality holds then one cannot expect the
boundary quarks to carry a definite spin. Rather they may be thought of in
terms of an infinite-dimensional reducible representation of the Poincare
group with any half-integer spin incorporated.
Of course when $F_{\mu\nu}$ reduces to the scalar case, the results of \cite{aabe} 
are fully reproduced.
These results are in excellent agreement with phenomenology.

We finally spelled out the restrictions that locality, chiral symmetry 
and conformal invariance place on
the couplings $F_{\mu\nu}$ and formulated the way to search for the consistent
realization of the $F_{\mu\nu}$ algebra.

\section*{Acknowledgments}
We are grateful to J. Alfaro for useful discussions and to P. Labra\~na for
 collaboration at the earlier stage of this work.
 A.A. was  supported by  Grant RFBR 05-02-17477 and Grant UR 02.01.299.
The work of D.E. and A.P. was supported by
the EURIDICE Network, by grant FPA2004-04582 and grant 2001SGR-00065.
Exchange visits were supported by the
CICYT-INFN bilateral agreements. A.P. acknowledges the support from a
graduate fellowship from Generalitat de
Catalunya.

\section*{Appendix A. Ein-bein projection of the Dirac operator on the
  string boundary}\label{Ap_einbein}
The Lagrangian \gl{lagmin} does not contain any operators that could
give rise to the anomalous P-odd part of the Chiral Dynamics.
To approach the required modification  let us guess on what might be the form of boundary
Lagrangian if one
derives it, say, from the essential part of the Chiral Quark Model projecting it on
the string boundary. In what follows the Minkowski space-time is
employed to keep the axial-vector vertex to be Hermitian.

Let us introduce the constituent quark fields to control properly the chiral symmetry
during the "ein-bein" projection,
\be
Q_L \equiv \xi^\dagger \psi_L,\qquad Q_R \equiv \xi\psi_R,\qquad  \xi^2\equiv U.
\la{chv}
\ee
Under chiral rotations $U \rightarrow \Omega_R U \Omega^+_{L}$ the fields $\xi$
transform as follows
\be
\xi \longrightarrow h_{\xi} \xi \Omega^+_{L} = \Omega_R \xi h^+_{\xi},
\la{nlr}
\ee
with $h_{\xi}$ being a nonlinear functional of fields $\xi$. As a consequence
the hidden vector
symmetry of the constituent field action replaces the original chiral invariance.

In these variables the CQM Lagrangian density and the pertinent E.o.M. read
\be
{\cal L}_{CQM} =
i \bar Q \left( \not\!\partial  +  \not\! v +
 g_A \not\! a \gamma_5\right) Q +\mbox{\rm mass terms};\quad  i \left( \not\!\partial  +  \not\! v +
 g_A \not\! a \gamma_5\right)Q +\mbox{\rm mass terms} = 0,
\la{CQM}
\ee
where
\ba
&& Q \equiv Q_L + Q_R,\qquad \not\!\! A \equiv \gamma^\mu  A_\mu,
\no
&& v_\mu \equiv \frac12(\xi^\dagger(\partial_\mu\xi) -
(\partial_\mu\xi) \xi^\dagger),
\qquad
a_\mu \equiv - \frac12 (\xi^\dagger(\partial_\mu\xi) + (\partial_\mu\xi) \xi^\dagger),
\la{va}
\ea
and $g_A \equiv 1 - \delta g_A$ is an axial coupling constant of
quarks to $\Pi$-ons.
We skip all mass effects of the CQM,
thereby neglecting the current quark mass in the chiral limit whereas relegating the
effects of constituent quark mass to the gluodynamics encoded in the string interaction.
Then one can decouple the left and right components of boundary fields in  the process
of dim-1 projection.

We assume the quark fields be located on the dim-1 boundary with coordinates
$x_\mu \equiv x_\mu(\tau)$. The  first step in
projection of the E.o.M. \gl{CQM} can be performed by
their multiplication on $\gamma^\mu \dot x_\mu$ which leads to the following boundary
equations,
\be
\left\{ i \left( \partial_\tau  +  \dot x_\mu v^\mu +
 g_A \gamma_5  \dot x_\mu a^\mu  \right)+ \sigma^{\mu\nu}\dot x_\mu
\left(\partial_\nu +  v_\nu +
 g_A \gamma_5 a_\nu  \right)
 \right\}Q =0;\qquad \sigma^{\mu\nu} \equiv \frac12 i [\gamma^\mu \gamma^\nu] .
\la{projec}
\ee
Notice that this projected Dirac-type equation seems to be associated to the boundary
action with a Lagrangian of type \gl{lagmin}. But in order to provide the correct Hermitian
properties of  the Lorentz symmetry generators $\sigma_{\mu\nu}$ one must involve
the Dirac conjugated spinors, $\bar\psi \equiv \psi^\dagger \gamma_0$. As a consequence,
the axial-vector part in the first, scalar contribution becomes anti-Hermitian as
$\gamma$ matrices anticommute. It can be cured by the prescription of analytic
continuation $g_A \rightarrow z = i g_A$ in the scalar part (only). At this place
we must adopt an arbitrary constant $z$ subject to the consistency conditions from the string
with boundary..

Let us restore the current quark basis of fields $\psi_L$ thereby going back to the original
chiral fields $U$. We use Eqs.\gl{chv} and multiply the left and right
component of Eq.\gl{projec} by
$\xi$ and $\xi^\dagger$ respectively. The result is that,
\ba
&&\frac12 \left\{ i\left( \{\partial_\tau, U^{\dagger}\}  +
 z  \dot U^{\dagger} \right)+ \sigma^{\mu\nu} \dot x_\mu
\left(\{\partial_\nu, U^{\dagger}\}+
 g_A  \partial_{\nu} U^{\dagger} \right)
 \right\}\psi_L =0;\no
&&\frac12 \left\{ i\left( \{\partial_\tau, U\}  +
 z  \dot U \right)+ \sigma^{\mu\nu} \dot x_\mu
\left(\{\partial_\nu, U\}+
 g_A  \partial_{\nu} U \right)
 \right\}\psi_R =0.
\la{projec1}
\ea
Now the culminating point of the "ein-bein" projection consists of making the
quark fields $\psi$ truly one-dimensional. Namely we define their gradient in terms of the tangent
vector $\dot x_\mu$ and  arbitrary matrix functions $f(x_\mu), b(x_\mu)$ of $x_\mu(\tau)$,
\ba
&&\partial_{\mu} (f_L\psi_L)+ \partial_{\mu} (b_L) \psi_L \Rightarrow \frac{\dot x_\mu}
{\dot x_\nu\dot x^\nu} \left[\partial_{\tau} (f_L\psi_L)+ \partial_{\tau} (b_L) \psi_L\right];\no
&&\partial_{\mu} (f_R\psi_R) + \partial_{\mu} (b_R) \psi_R\Rightarrow \frac{\dot x_\mu}
{\dot x_\nu\dot x^\nu} \left[\partial_{\tau} (f_R\psi_R)+ \partial_{\tau} (b_R) \psi_R\right].
\ea
Keeping in mind our program we choose the functions $f(x_\mu), b(x_\mu)$ to provide the correct
chiral properties, translational and reparameterization invariance (in terms of chiral
fields $U$). As well the operator appeared in projection must be anti-self-adjoint in respect to
the dim-4 Dirac scalar product. All these requirements are satisfied by the
choice,
\be
\{\partial_\mu, U^{\dagger}\}\psi_L \Rightarrow \frac{\dot x_\mu}{\dot x_\nu\dot x^\nu}
\{\partial_\tau, U^{\dagger}\}\psi_L;\qquad
\{\partial_\mu, U\}\psi_R \Rightarrow \frac{\dot x_\mu}{\dot x_\nu\dot x^\nu}
\{\partial_\tau, U\}\psi_R.
\ee
Finally, the projected equations are originated from the boundary Lagrangian,
\ba
L^{(f)}&=&\frac12  i \left\{\bar\psi_L \left[ \{\partial_\tau, U\}  +
 z \dot U + g_\sigma \sigma^{\mu\nu} \dot x_\mu
   \partial_{\nu} U\right] \psi_R  + \bar\psi_R \left[ \{\partial_\tau, U^{\dagger}\}  -
 z^* \dot U^{\dagger} - g^*_\sigma \sigma^{\mu\nu} \dot x_\mu
   \partial_{\nu} U^{\dagger}\right]\psi_L \right\};\no
   &\equiv& \frac12 i  \left\{\bar\psi_L \left[ \{\partial_\tau, U\}  + 
\widehat{F}^{\mu\nu} \dot x_\mu
   \partial_{\nu} U\right] \psi_R  + \bar\psi_R \left[ \{\partial_\tau, U^{\dagger}\} -
\widehat{F}_\sharp^{\mu\nu} \dot x_\mu
   \partial_{\nu} U^{\dagger}\right]\psi_L \right\};\no
   \widehat{F}^{\mu\nu} &\equiv& z g^{\mu\nu} + g_\sigma \sigma^{\mu\nu};\quad
  \widehat{F}_\sharp^{\mu\nu} \equiv  \gamma_{0}\left(\widehat{F}^{\mu\nu}\right)^{\dagger}\gamma_{0},
\label{lagfull}
\ea
where we have obtained the indications that $g_\sigma = -i g_A$. Still
keeping
in mind a certain ambiguity in the projection procedure we must
consider  both constants $z$ and $g_\sigma$
as arbitrary ones and search for their values from the consistency of the hadron string with
chiral fields on its boundary.

The meaning of purely imaginary $z$ and $g_\sigma$ is clarified by the
CP symmetry \gl{CP} of the Lagrangian \gl{lagfull}. Indeed it is CP
symmetric only if
\be
z = - z^*;\quad g_\sigma = - g_\sigma^*. \label{CPfull1}
\ee

\newpage

\section*{Appendix B. One-loop two-fermion one-boson vertex}

In this Appendix we present the calculation of 1-boson vertex for the
boundary Lagrangian (\ref{Lagran4D}) including a more general spin structure
$F_{\mu\nu}$.

\begin{center}
  \begin{tabular}{|c|c|c|}
    \hline
    Coefficient & U structure & F structure\\
    \hline
 \multicolumn{3}{|c|}{}\\
    \multicolumn{3}{|c|}{Divergent part}\\
 \multicolumn{3}{|c|}{}\\
    \hline
    $- \frac{1}{4} \theta ( A - B ) \Delta ( 0 )$ & $U^{\dag} \partial_{\eta}
    \partial_{\sigma} \partial_{\lambda} U \nonesep U^{\dag}$ &
    $\delta_{\sigma \lambda} [ \bar{X}_{\rho} ( A ) ( \delta_{\eta \rho} +
    F_{\eta \rho} ) + \bar{X}_{\rho} ( B ) ( \delta_{\eta \rho} - F_{\eta
    \rho} ) ]$\\
    \hline
    $\frac{1}{4} \theta ( A - B ) \Delta ( 0 )$ & $U^{\dag} \partial_{\eta}
    \partial_{\sigma} U \nonesep U^{\dag} \partial_{\lambda} U \nonesep
    U^{\dag}$ & $[ \bar{X}_{\rho} ( A ) ( \delta_{\sigma \lambda} + F_{\eta
    \rho} ) ( \delta_{\sigma \lambda} + F_{\lambda \sigma} ) + \bar{X}_{\rho}
    ( B ) ( 2 \delta_{\eta \rho} \delta_{\sigma \lambda}$\\
    &  & $- F_{\eta \rho} \delta_{\sigma \lambda} - \delta_{\eta \rho}
    F_{\lambda \sigma} + \delta_{\eta \rho} F_{\sigma \gamma} F_{\lambda
    \gamma} - F_{\eta \rho} F_{\lambda \sigma} ) ]$\\
    \hline
    $\frac{1}{4} \theta ( A - B ) \Delta ( 0 )$ & $U^{\dag} \partial_{\sigma}
    U \nonesep U^{\dag} \partial_{\lambda} \partial_{\eta} U \nonesep
    U^{\dag}$ & $[ \bar{X}_{\rho} ( B ) ( \delta_{\sigma \lambda} - F_{\sigma
    \lambda} ) ( \delta_{\sigma \lambda} - F_{\eta \rho} ) + \bar{X}_{\rho} (
    A ) ( 2 \delta_{\eta \rho} \delta_{\sigma \lambda}$\\
    &  & $- F_{\eta \rho} \delta_{\sigma \lambda} - \delta_{\eta \rho}
    F_{\sigma \lambda} + \delta_{\eta \rho} F_{\sigma \gamma} F_{\lambda
    \gamma} - F_{\sigma \lambda} F_{\eta \rho} ) ]$\\
    \hline
    $- \frac{1}{8} \theta ( A - B ) \Delta ( 0 )$ & $U^{\dag}
    \partial_{\sigma} U \nonesep U^{\dag} \partial_{\eta} U \nonesep U^{\dag}
    \partial_{\lambda} U \nonesep U^{\dag}$ & $[ \bar{X}_{\rho} ( A ) ( ( 3
    \delta_{\eta \rho} + F_{\eta \rho} ) ( \delta_{\sigma \lambda} +
    F_{\lambda \sigma} )$\\
    &  & $+ F_{\sigma \gamma} ( \delta_{\eta \rho} - F_{\eta \rho} ) (
    \delta_{\lambda \gamma} + F_{\lambda \gamma} ) )$\\
    &  & $+ \bar{X}_{\rho} ( B ) ( ( \delta_{\sigma \lambda} - F_{\sigma
    \lambda} ) ( 3 \delta_{\eta \rho} - F_{\eta \rho} )$\\
    &  & $- ( \delta_{\sigma \gamma} - F_{\sigma \gamma} ) ( \delta_{\eta
    \rho} + F_{\eta \rho} ) F_{\lambda \gamma} ) ]$\\
    \hline
 \multicolumn{3}{|c|}{}\\
    \multicolumn{3}{|c|}{Finite part $\propto \Delta ( A - B )$}\\
 \multicolumn{3}{|c|}{}\\
    \hline
    $\frac{1}{4} \theta ( A - B ) \Delta ( A - B )$ & $U^{\dag}
    \partial_{\eta} \partial_{\sigma} U \nonesep U^{\dag} \partial_{\lambda} U
    \nonesep U^{\dag}$ & $\bar{X}_{\rho} ( A ) \delta_{\eta \rho} (
    \frac{1}{2} \delta_{\sigma \lambda} + F_{\sigma \lambda} - F_{\lambda
    \sigma} - F_{\sigma \gamma} F_{\lambda \gamma} )$\\
    \hline
    $\frac{1}{4} \theta ( A - B ) \Delta ( A - B )$ & $U^{\dag}
    \partial_{\sigma} U \nonesep U^{\dag} \partial_{\lambda} \partial_{\eta} U
    \nonesep U^{\dag}$ & $\bar{X}_{\rho} ( B ) \delta_{\eta \rho} (
    \frac{1}{2} \delta_{\sigma \lambda} - F_{\sigma \lambda} + F_{\lambda
    \sigma} - F_{\sigma \gamma} F_{\lambda \gamma} )$\\
    \hline
    $- \frac{1}{8} \theta ( A - B ) \Delta ( A - B )$ & $U^{\dag}
    \partial_{\sigma} U \nonesep U^{\dag} \partial_{\eta} U \nonesep U^{\dag}
    \partial_{\lambda} U \nonesep U^{\dag}$ & $[ \bar{X}_{\rho} ( A ) (
    \delta_{\sigma \gamma} + F_{\sigma \gamma} ) ( \delta_{\eta \rho} +
    F_{\eta \rho} ) ( \delta_{\lambda \gamma} - F_{\lambda \gamma} )$\\
    &  & $+ \bar{X}_{\rho} ( B ) ( \delta_{\sigma \gamma} + F_{\sigma \gamma}
    ) ( \delta_{\eta \rho} - F_{\eta \rho} ) ( \delta_{\lambda \gamma} -
    F_{\lambda \gamma} ) ]$\\
    \hline
 \multicolumn{3}{|c|}{}\\
    \multicolumn{3}{|c|}{Finite part $\propto \int\limits_B^A \mathd \tau \dot{\bar{X}}_{\rho} ( \tau )
    \Delta ( A - \tau ) \bignone \equiv \tmop{Int}_A$}\\
 \multicolumn{3}{|c|}{}\\
    \hline
    $\frac{1}{4} \theta ( A - B ) \tmop{Int}_A$ & $U^{\dag} \partial_{\sigma}
    U \nonesep U^{\dag} \partial_{\lambda} \partial_{\eta} U \nonesep
    U^{\dag}$ & $( \delta_{\sigma \gamma} + F_{\sigma \gamma} ) (
    \delta_{\lambda \gamma} F_{\eta \rho} - \delta_{\eta \rho} F_{\lambda
    \gamma} )$\\
    \hline
    $- \frac{1}{8} \theta ( A - B ) \tmop{Int}_A$ & $U^{\dag}
    \partial_{\sigma} U \nonesep U^{\dag} \partial_{\eta} U \nonesep U^{\dag}
    \partial_{\lambda} U \nonesep U^{\dag}$ & $( \delta_{\sigma \gamma} +
    F_{\sigma \gamma} ) ( \delta_{\eta \rho} - F_{\eta \rho} ) (
    \delta_{\lambda \gamma} + F_{\lambda \gamma} )$\\
    \hline
 \multicolumn{3}{|c|}{}\\
    \multicolumn{3}{|c|}{Finite part $\propto \int\limits_B^A \mathd \tau \dot{\bar{X}}_{\rho} ( \tau )
    \Delta ( \tau - B ) \bignone \equiv \tmop{Int}_B$}\\
 \multicolumn{3}{|c|}{}\\
    \hline
    $\frac{1}{4} \theta ( A - B ) \tmop{Int}_B$ & $U^{\dag} \partial_{\eta}
    \partial_{\sigma} U \nonesep U^{\dag} \partial_{\lambda} U \nonesep
    U^{\dag}$ & $( \delta_{\sigma \gamma} F_{\eta \rho} - F_{\sigma \gamma}
    \delta_{\eta \rho} ) ( \delta_{\lambda \gamma} - F_{\lambda \gamma} )$\\
    \hline
    $\frac{1}{8} \theta ( A - B ) \tmop{Int}_B$ & $U^{\dag} \partial_{\sigma}
    U \nonesep U^{\dag} \partial_{\eta} U \nonesep U^{\dag} \partial_{\lambda}
    U \nonesep U^{\dag}$ & $( \delta_{\sigma \gamma} - F_{\sigma \gamma} ) (
    \delta_{\eta \rho} + F_{\eta \rho} ) ( \delta_{\lambda \gamma} -
    F_{\lambda \gamma} )$\\
    \hline
  \end{tabular}

\end{center}

\end{document}